\documentclass[letterpaper,twocolumn,10pt]{article}
\usepackage{usenix2019_v3}
\usepackage[utf8]{inputenc}\usepackage[T1]{fontenc}
\usepackage{graphicx,booktabs,amsmath,amssymb,url,microtype,array,upquote,placeins}
\hypersetup{hidelinks}
\pdftrailerid{}
\begin{document}
\title{\Large \bf Where the Numbers Come From: Auditing Evaluation in Provenance-Based Intrusion Detection}
\author{%
{\rm Jihwan Moon}\\
INFOCZ Inc \\
\texttt{jihwan@infocz.co.kr}
\and
{\rm Gunhee Kim}\\
Seoul National University \\
\texttt{gunhee@snu.ac.kr}
\and
{\rm Myeongjang Pyeon}\\
INFOCZ Inc \\
\texttt{pmaster@infocz.co.kr}}
\maketitle
\begin{abstract}

Reproducing a provenance-based intrusion detector's score does not establish what that score says
about its emitted alarms or the information its encoder uses. We audit nine released implementations,
execute four detectors using their own code, and isolate three measurement effects. First, a fixed-alert
comparison separates label choice from neighbourhood credit: ThreaTrace reports precision 0.938 with neighbourhood credit, although
only seven of its 994 alarms carry its own attack label. Second, removing test-label checkpoint selection
lowers attack detection precision (ADP) by 0.16 to 0.33 across four forty-member word2vec configurations without
changing detector order. Third, a buffer-reuse defect gives a linear encoder unintended degree-dependent
inputs. Correcting it lowers type-only ADP in every identical-input initialization pair on two hosts,
while historical word2vec effects depend on the host. These findings qualify claims from the inspected implementations
about alarm precision, performance magnitude and static-attribute sufficiency. STRICT connects them to six checkable reporting
requirements. Because the comparisons condition on benchmark targets, they neither validate those labels nor
establish a universal detector ranking.

\end{abstract}
\section{Introduction}

Provenance-based intrusion detection systems (PIDS) model processes, files, sockets, and their interactions.
Recent systems report near-perfect node- or window-level detection scores on large provenance graphs\cite{cheng2024kairos,rehman2024flash,jia2024magic,li2024nodlink,jiang2025orthrus}, but a score alone does not tell an
operator how often the detector alerts, how quickly it detects a campaign, or whether its threshold transfers to a
longer observation period. For an operator, a model that ranks attacks well but crosses its threshold thousands of
times is not equivalent to one that raises a few alerts.

Bilot et al. already demonstrate problems with granularity, thresholds and initialization variability
in an eight-system comparison\cite{bilot2025velox}, and PIDSMaker provides a common experimental implementation
\cite{bilot2026pidsmaker}. Guerra et al. study benchmark quality and architectural conclusions under shared labels
and validation-based selection\cite{guerra2026benchmarks}. Building on these studies, we ask
which conclusions does a reproduced result actually support?

Three distinctions matter. Precision over neighbourhood-credited detections need not describe
the alarms a detector emitted. Agreement on the winning detector need not validate the reported
performance magnitude. A linear encoder need not receive only static attributes. We test these
distinctions by holding alerts, trained members or realized inputs fixed and changing the operation
under examination. This separates a change in measurement or information access from a change in
detector capability. Our contributions are:

\begin{enumerate}
\item \textbf{Reported precision versus emitted alarms.} A complete label-by-scorer comparison on
unchanged ThreaTrace alarms isolates neighbourhood credit from the supplied labels (\S3). High credited
precision coexists with few labelled emitted alarms even under the detector's own target list.
The label contrast also changes direction with the scorer. Harmonizing labels alone therefore
does not make reported precisions comparable.

\item \textbf{Stable ordering versus test-independent performance.} Source inspection identifies direct test-label selection
in four released implementations. Reselecting checkpoints on the same trained members quantifies the
effect in PIDSMaker (\S3.4). The headline word2vec ordering survives, but reported performance falls.
A cross-member diagnostic attributes most of the gap to member-specific checkpoint choice rather
than a generally better earlier epoch. Ranking agreement alone misses this dependence.

\item \textbf{A linear encoder versus static-attribute sufficiency.} We trace degree-dependent feature scaling to
the original Velox release and PIDSMaker, then test the correction using paired retraining and
identical-input controls (\S3.6, \S5.2). The released encoder receives information beyond its declared
static attributes. Non-GNN competitiveness can survive while its attribution to static attributes
does not follow from that comparison.

\end{enumerate}
STRICT turns these findings into six reporting requirements (\S4), whose principles have established
antecedents. Its role is to make the checks needed for each conclusion explicit: identify what the scorer
counts, establish what information selects the reported model, and verify what inputs the encoder
actually receives. The contribution is the controlled evidence for these distinctions, not another
universal detector ordering. Identical-input controls isolate the feature-processing effect more
strongly than historical seed matching, although neither design establishes which target labels are correct.

The source audit covers nine released implementations, of which
Flash, MAGIC, ThreaTrace and CAPTAIN are also executed with their own code. The broader PIDSMaker comparison
covers nine integrated families on multiple benchmark hosts. It describes specified configurations, not
additional independent replications of the three mechanisms or a universal detector ranking.
Supporting analyses distinguish ranking from fixed-budget campaign coverage and chronological alert
delivery. Their purpose is to clarify the measured quantity, not to establish deployment readiness.

\section{Background, threat model, and setup}

\textbf{Definitions.} A \emph{member} is one trained model of a configuration under specified random seeds,
and $n$ is the member count. \emph{Campaign coverage} is the fraction of labelled attack campaigns with at least one true
positive. \emph{ADP} (attack detection precision) is PIDSMaker's ranking summary: walking the scores in descending order, it
takes the highest campaign coverage reached at each precision and integrates it over precision. It is \emph{tie-safe} when
nodes sharing a score enter together.

ADP summarizes campaign coverage across precision values, not across alert counts. It does not identify
which detector reaches every campaign within a fixed ranking budget (\S3.4). We report it because PIDSMaker uses it to rank systems and select checkpoints.
Section 3.4 therefore measures selection in the procedure's own units. Like any precision-based score, ADP inherits its label set.
It describes retrospective ranking, whereas the chronological replay of \S3.5 measures alert workload.

A metric is \emph{strict} when a node counts as detected only if it is itself alerted and labelled.
The \emph{2-hop relaxation}, or neighbourhood credit, used by Flash and ThreaTrace credits detections and
drops false positives within two hops of a labelled node.

\textbf{Precision across members.} Under strict scoring, pooled precision weights members by
their alert counts, whereas mean member precision weights members with defined precision equally.
They differ when alert volumes vary, and a silent member has undefined precision.

The \emph{max-of-validation} operating point, PIDSMaker's native threshold,
alerts on every test node whose score exceeds the maximum score of the validation period. For this rule,
\emph{generation consistency} requires saved alerts to agree with threshold decisions reconstructed from test scores
and from validation node scores, where a node's validation score is its maximum loss among the validation losses
recorded by the member's training execution. This is a necessary
consistency check, not proof of training provenance. A mismatched
record cannot contribute to a comparison. Replacing it requires a documented regeneration and a new evaluation
record, not adjustment of its saved alerts. Detectors with other threshold rules require their own checks.

An \emph{offline ranking budget} $B$ selects the top-$B$ test nodes over the complete test period, ties broken
by node id. Precision at this budget is $P@B=\mathrm{TP}(B)/B$. It measures retrospective ranking, not a causal alert limit per day: future scores can determine
which earlier nodes enter the set. The chronological workload analysis in \S3.5 separately replays alert policies
on saved scores of fixed two-hour blocks. $B^{*}$ is the smallest budget at which every labelled campaign has a true positive. It is computed
with test labels after evaluation, not used to choose a deployable operating point.

\textbf{Comparison conditions.} A measurement specifies the evaluated entities and period, label set,
scoring rule, model-selection rule and alert policy. Comparisons of the same quantity require a common entity
population, observation period, target labels and metric. Preprocessing, selection rules and alert policies
must either be held fixed or treated explicitly as part of the compared detector procedures. An evaluation
contrast changes a declared condition while retaining the same predictions or trained
members. When labels or scoring rules change, the contrast measures sensitivity of the reported quantity,
not an improvement or deterioration of the detector. We retain a detector's native evaluation unit when no
validated mapping to another unit exists. Node precision and window precision therefore cannot establish a
cross-system ranking merely because both are called precision.

These conditions distinguish three questions. Offline ranking asks which entities receive high
scores in a completed period. Chronological alerting asks which campaigns a policy identifies using only
information available at each decision time. Deployment utility additionally requires evidence about the
correctness and cost of analyst decisions. We measure ranking and chronological policy replay on the stated hosts.
The replay assumes block scores are available at block end and does not independently establish causal detector inputs. Neither node-level
precision nor alert counts establish the third. Training repetitions quantify variability conditional on
those hosts, not uncertainty over a population of future deployments.

\textbf{Primary evaluation and test-selected diagnostic.} Our \emph{STRICT evaluation} groups tied scores
and selects checkpoints without test attack labels. The final checkpoint is the primary retrospective rule,
with minimum benign-validation loss as a sensitivity analysis. The \emph{test-ADP reconstruction} selects
the largest recorded test ADP at its recorded precision, taking the first recorded checkpoint on a tie.
Native PIDSMaker instead breaks equal ADP by discrimination, so this diagnostic is not an exact native
selector reproduction. We quantify the headline sensitivity to that difference in \S3.4.
We recompute tie-safe ADP for the selected output, which need not maximize that reported metric.
Paired comparisons hold the available trained members fixed while changing
the specified evaluation rule. Results from defective feature processing remain implementation diagnostics.
To assess that processing, we use separately trained corrected controls (\S5).

A \emph{member vote} alerts on a node when at least $k$ of $n$ members alert on it, each at its own threshold.
Section 3.5 evaluates a vote under padding, and a separate preregistered test evaluates it on held-out hosts.
Each comparison states its member count and randomization source. The full configuration inventory
remains in the retained experimental record.

Velox\cite{bilot2025velox} and Orthrus\cite{jiang2025orthrus} share PIDSMaker\cite{bilot2026pidsmaker}'s benign-only edge-type prediction objective and its node-level evaluation.
Velox applies a linear encoder to node attributes, whereas Orthrus adds temporal memory and graph attention. For
each node, the procedure reduces the losses of its incident edges to a maximum score and alerts when that score
exceeds the maximum validation score. We compare word2vec\cite{mikolov2013word2vec}, hierarchical hashed-path (HFH),
and node-type-only features. Each table states the number of recoverable members.
Headline word2vec configurations vary featurization seeds at a fixed training seed, whereas deterministic-feature
configurations vary training seeds. The identical-input controls instead vary initialization between
pairs while fixing prepared inputs. Member intervals describe these randomization schemes on fixed hosts.

Cross-system comparisons condition on an author-assembled frozen roster, not a complete or
prospectively selected training search. Overlapping expansions under the same settings retain the larger
configuration and exclude the smaller as a whole. Because label variants can reuse trained outputs,
evaluation records are not pooled as independent training repetitions.

Evaluation covers DARPA E3\cite{darpa2020tc} CADETS, THEIA and CLEARSCOPE, OpTC\cite{darpa2020optc} hosts 201, 501 and 051,
and E5 CADETS, THEIA and CLEARSCOPE. Common-implementation comparisons also use two endpoint detection and response (EDR) datasets, ATLASv2\cite{riddle2024atlasv2} and Carbanak v2\cite{liu2026edrprov}. E3 FIVEDIRECTIONS and TRACE are held out for one preregistered test, while CADETS E5 supplies a separate preregistered capacity test (\S3.5). We report TP,
FP, precision, campaign coverage, and tie-safe ADP. Temporal analyses also report delay and alerts per observed host-day, with each DARPA dataset here representing one host and OpTC's hosts counted separately. Under its configured dates CADETS E3 supports offline ranking only, while
THEIA E3/E5 and CADETS E5 support chronological analyses.

The defender observes host provenance events and trains only on a designated benign period. The attacker
may choose processes, files, and network actions during the later test period but cannot alter training code or
ground-truth files. We do not assume the labels are complete, so we report attack-adjacent analysis only as
false-positive forensics. Poisoning, sensor compromise, and cross-host correlation remain out of scope.

The padding experiment instantiates a narrower attacker: it uses campaign identities and windows to
attach additional events to attack-descendant processes and catalogue files, but observes neither detector
scores nor gradients.
It tests alert-volume and capacity effects, not detector-aware evasion. An alert on an original labelled node
after padding does not establish that the detector identified the original malicious event (\S3.5).

\section{Where the numbers come from}
\begin{table}[t]\centering\setlength{\tabcolsep}{4pt}\small
\begin{tabular}{lrr}
\toprule
Labels & Strict & Neighbourhood credit \\
\midrule
ThreaTrace & $7/994=0.0070$ & $12847/13690=0.9384$ \\
Orthrus & $9/994=0.0091$ & $26/704=0.0369$ \\
\bottomrule
\end{tabular}
\caption{ThreaTrace on CADETS: 994 unchanged alarms among 357,173 scored nodes, with 12,852 ThreaTrace targets and 46 Orthrus targets. Cells give counted fraction and precision. Neighbourhood credit changes both numerator and denominator, so it is not precision over emitted alarms.}\label{tab:native-factorial}
\end{table}

A reported precision can change even when no alert changes. We begin with this fixed-alert
comparison, then examine checkpoint selection (\S3.1, \S3.4) and chronological workload (\S3.5). Feature and
false-positive diagnostics explain limits of these comparisons without establishing an architecture
advantage or an alternative detection metric.

\textbf{Labels and scoring on fixed alerts.} Changing only the label set can change precision
on unchanged alarms. MAGIC's released CADETS output has precision 0.944 with ThreaTrace labels but
below 0.001 with Orthrus labels. Its THEIA output shows the same direction.
The threshold is fixed and test-selected. This contrast isolates label sensitivity, not which
labels are correct or the effect of choosing that threshold.

ThreaTrace's 994 CADETS alarms yield the complete two-by-two comparison in Table~\ref{tab:native-factorial}.
With its own labels, neighbourhood credit reports precision 0.938, although only seven emitted alarms are
labelled. Let $P_{L,g}$ denote precision under labels $L$ and scorer $g$. Replacing ThreaTrace labels ($T$)
with Orthrus labels ($O$) increases strict precision but decreases neighbourhood-credit precision.
For neighbourhood credit $r$ and strict scoring $s$, the interaction is
$\Delta=(P_{O,r}-P_{O,s})-(P_{T,r}-P_{T,s})=-0.9035$.

Because the label contrast changes sign with the scorer, shared labels alone do not make two reported
precisions comparable. The relaxed scorer also changes the denominator, so its precision is not the fraction
of emitted alerts that are labelled. Holding ThreaTrace's own target list fixed establishes this difference
in counted outcomes independently of which alternative target is more accurate. This is an exact contrast
for one saved prediction set, not a population effect or a judgment about which labels are correct.

Entity-level replay reproduces all eight native scoring cells from retained scores, mappings and targets.
Checks of MAGIC's retained CADETS graph and score-cache identity establish
consistency, not fresh inference or semantic validation.

Much of MAGIC's precision difference reflects the label sets' different prevalence.
Normalizing precision by that prevalence separates the base-rate effect from alert concentration.
ThreaTrace's strict-alert concentration changes in the opposite direction, consistent with the
factorial contrast. Neither comparison adjudicates the target lists.

\subsection{Test-selected epochs and seeds}
\begin{figure*}[t]\centering\includegraphics[width=0.95\textwidth]{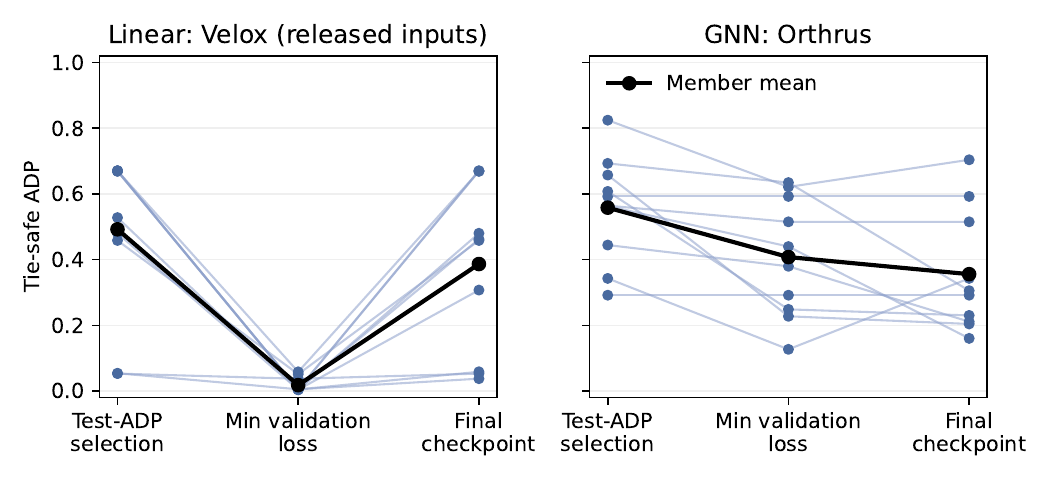}\caption{Checkpoint choice changes the apparent encoder gap. Each thin line joins the same trained member under three selectors, and black lines show means (ten training-seed members per encoder, CADETS E3, type-only features). Test-ADP selection uses the reconstruction of \S2, while the other rules use no test attack labels. The linear encoder retains unintended degree-dependent inputs, so this illustrates the released implementation, not a clean architecture comparison. The small final mean-order change is descriptive, not a resolved rank reversal.}\label{fig:rules}\end{figure*}

Checkpoint selection changes which output is reported, even when the trained members and metric
are fixed. We take the final saved epoch as the primary retrospective rule, because it performs no checkpoint
selection, and retain minimum benign-validation loss as a complete sensitivity analysis. Neither rule
guarantees a good detector: minimum validation loss can select the first evaluated checkpoint.
The paired estimates in \S3.4 measure the selection effect, whereas best-member reports do not estimate mean performance.

Figure~\ref{fig:rules} makes this dependence visible on CADETS type-only features. The same ten members
of each encoder show a large GNN advantage under minimum benign-validation loss, but much closer means
under test selection and the final checkpoint. The small final-checkpoint mean difference favours the
linear encoder but does not establish a reliable ordering. Six linear members select the first evaluated
checkpoint under benign-loss selection. This illustrates how a rule can interact with model training,
not that one label-free rule is universally preferable. The linear members use the released reindexer,
so this is an implementation diagnostic, separate from the corrected word2vec comparison in \S3.4.

\textbf{Checkpoint experiment.} All 160 members in the four headline word2vec configurations complete
12 training passes. The seven evaluated checkpoints have zero-based indices $0,1,3,5,7,9,11$.
Epoch 0 follows the first training pass, not untrained initialization. Final selection takes epoch 11.
The other selectors choose among the same seven outputs using recorded mean benign-validation edge loss
or test ADP, retaining the recorded precision and tie conventions (\S2). This holds both
the trained members and the checkpoint search opportunities fixed. We use percentile 95 \% intervals
from 2,000 member resamples. For paired contrasts, we resample within-member differences, not checkpoint outputs independently.

\textbf{Member-specific choice versus a shared epoch.} To check whether a generally better earlier
epoch explains the gap, we choose each member's checkpoint using the other 39 members' mean tie-safe ADP,
breaking ties toward the earlier epoch. The cross-member column in Table~\ref{tab:checkpoint-core} reports this
diagnostic alongside the three selectors. Across the four configurations, the test-ADP reconstruction
exceeds this cross-member choice by 0.12--0.33 mean ADP, whereas cross-member and final means differ by
at most 0.04 in absolute value. The larger component is therefore member-specific checkpoint choice,
not a shared late-epoch decline. Because this diagnostic still uses the same host's test labels, it estimates
neither test-independent performance nor optimism on new hosts.

A source audit finds direct test-label selection in Orthrus, Velox, ThreaTrace and MAGIC.
Orthrus and Velox select checkpoints using test metrics. ThreaTrace accepts a model cascade using
test recall and precision, while MAGIC selects a threshold from its test precision--recall curve.
PIDSMaker's shared test-ADP selector extends this dependence to its integrated families.
The inspected revisions and decisive source excerpts are in Table~\ref{tab:anchors} and Table~\ref{tab:excerpts}.
These findings concern the inspected implementations, not the prevalence of leakage across the field.

MAGIC's released checkpoint reproduces its reported CADETS result under ThreaTrace labels.
Its paper describes a benign false-positive-rate threshold, whereas the released evaluator uses a
test-recall target. Our fixed-alert comparisons hold that threshold fixed and do not measure
the effect of replacing it with a benign-only threshold.

MAGIC's training data also include events recorded after the test window, and its preprocessing removes
training edges incident to labelled nodes. Its released procedure is therefore neither chronological nor
independent of attack labels. Orthrus also uses test data, but in a different way:
it replaces its validation threshold using the unlabelled test-score distribution. This is test-distribution
adaptation, not direct test-label selection.

By contrast, Flash\cite{rehman2024flash}, NodLink\cite{li2024nodlink}, CAPTAIN
\cite{wang2025captain}, and Reha et al.'s implementation expose fixed, training-derived, or validation-derived
selection rules without direct attack-label contact. Kairos\cite{cheng2024kairos} uses hard-coded per-day queue
cutoffs of undocumented provenance, which we classify as opaque rather than as presumed leakage.

\subsection{Thresholds are sample-size-dependent order statistics}
\begin{figure*}[t]\centering\includegraphics[width=0.86\textwidth]{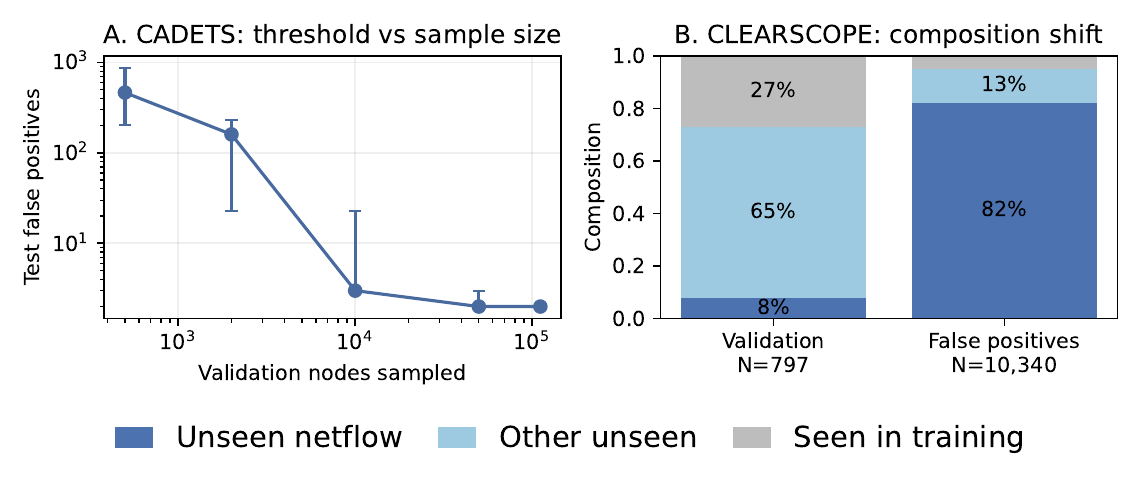}\caption{A benign-validation threshold need not imply stable false-alarm workload. (A) Fixed CADETS model and test scores: median and interquartile range over 100 validation subsamples per reduced size. The full validation set is evaluated once. (B) A separate CLEARSCOPE model: composition of validation nodes and false-positive nodes. Unseen means absent from training. Both use final-checkpoint member 1 and the maximum validation score. These are single-member calibration illustrations, not cross-host performance estimates.}\label{fig:valsize}\end{figure*}

The maximum validation score is an extreme order statistic, so it depends on how much validation data there
is. Even when benign validation and test scores are exchangeable and untied, a benign test node exceeds the maximum
of $N_{\mathrm{val}}$ validation scores with probability $1/(N_{\mathrm{val}}+1)$. The false-alarm rate is therefore set
by the validation size rather than chosen, and the expected number of false positives among $N_{\mathrm{test}}$ benign test
nodes is $N_{\mathrm{test}}/(N_{\mathrm{val}}+1)$. Conformal evaluation of malware classifiers calibrates its thresholds on
held-out scores under the same exchangeability, but chooses its rejection thresholds rather than inheriting
$1/(N_{\mathrm{val}}+1)$, and treats departures from exchangeability as concept drift\cite{jordaney2017transcend,barbero2022transcendent}.

Figure~\ref{fig:valsize} separates two calibration sensitivities. Subsampling one CADETS member's
validation day raises median false positives from 2 to 464 without retraining or changing its test scores.
On CLEARSCOPE, unseen netflows dominate a different member's false alerts. These illustrate sensitivity
to calibration size and population, not an estimate of how often either mechanism occurs.

Node scores are maxima over windows, so a threshold fitted to a short period need not transfer
to a longer one. The validation-size diagnostic does not isolate horizon length from distribution
change or rule out other benign-only calibration methods.

\subsection{False-positive forensics}

Attack proximity alone does not establish an alert's usefulness. Corrected Velox's false
positives and type- and degree-matched benign controls are both frequently close to labelled nodes,
although their fractions need not be equal. We treat adjacency as forensics, not detection credit.

\subsection{Checkpoint selection and fixed-budget rankings}
\begin{figure*}[t]\centering\includegraphics[width=\textwidth]{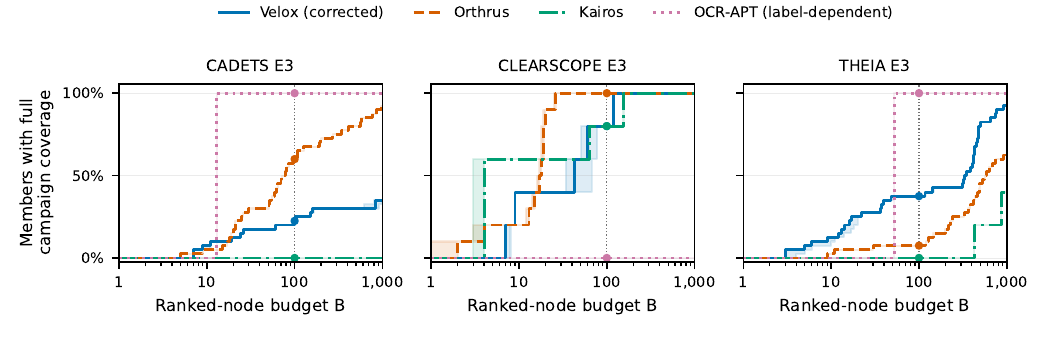}\caption{Full-campaign coverage versus ranked-node budget, final checkpoint. Each curve gives the fraction of recorded members with $B^{*}\le B$, not mean campaign coverage. Dots mark the previously declared budget $B=100$. Shading bounds every within-score tie order, not statistical uncertainty. The same scored nodes and labels are used within each host. Configurations have five to forty members. The explicitly labelled OCR-APT curve is a label-dependent supervision diagnostic (\S3.4), not a fair unsupervised-training comparison. Repeated outcomes are not independent confirmations. This post-hoc comparison concerns whole-test offline rankings, not daily alert acceptance.}\label{fig:coverage-budget}\end{figure*}

\begin{table*}[t]\centering\setlength{\tabcolsep}{4pt}\small
\begin{tabular}{lrrrrr}
\toprule
Model & Test-selected & Cross-member & Final & Benign-val. & Paired decrease [95 \% CI] \\
\midrule
\addlinespace[2pt]\multicolumn{6}{l}{\textbf{CADETS E3}} \\
Orthrus (W2V, GNN) & 0.526 & 0.257 & 0.280 & 0.297 & 0.245 [0.178, 0.323] \\
Velox (W2V, linear) & 0.334 & 0.217 & 0.179 & 0.168 & 0.155 [0.108, 0.201] \\
\addlinespace[2pt]\multicolumn{6}{l}{\textbf{THEIA E3}} \\
Orthrus (W2V, GNN) & 0.565 & 0.311 & 0.274 & 0.277 & 0.292 [0.218, 0.369] \\
Velox (W2V, linear) & 0.755 & 0.423 & 0.423 & 0.447 & 0.331 [0.239, 0.426] \\
\bottomrule
\end{tabular}
\caption{Mean tie-safe ADP on the same forty trained word2vec members per configuration. Test-selected uses the recorded-ADP reconstruction (\S2). Cross-member selects each member's epoch by the other 39 members' mean test ADP, with earliest-epoch tie-breaking. Both use test labels. Final and minimum benign-validation loss (benign-val.) do not. The paired decrease is test-selected minus final, with a 95 \% member-bootstrap interval. Within-host ordering is unchanged across the selectors. Velox uses the corrected reindexer.}\label{tab:checkpoint-core}
\end{table*}

The paired comparison shows checkpoint sensitivity without a robust rank reversal
(Table~\ref{tab:checkpoint-core}). Replacing test-label selection with the final checkpoint lowers mean ADP
by 0.16--0.33 on the same trained members. Orthrus remains above corrected word2vec Velox on
CADETS, and below it on THEIA. This in-sample selection contrast does not estimate performance loss
on new hosts. A change in architecture order
is neither required nor observed in these four configurations. Applying native discrimination tie-breaking
leaves the reported headline paired effects unchanged.

\textbf{Beyond two systems.} Across all 30 main-table configurations, final-checkpoint mean ADP is
lower in 26 than under recorded test-ADP selection, with a median decrease across configurations of 0.032.
There is no pooled rank-agreement statistic: R-CAID scores different node populations on all three hosts,
omitting 25 of the 118 positive targets on THEIA. Within-configuration checkpoint contrasts still hold
the scored population fixed. A common implementation alone does not ensure comparable detector rankings.

The broader nine-family comparison retains Orthrus labels, and its cross-label ADP ordering has not
been tested. ThreaTrace's lists supply no campaign identities for the same ADP or $B^{*}$ comparison.
Section 5.1 instead tests node-level sensitivity of the headline outputs within their fixed scored populations.
Neither analysis establishes which labels describe detection effectiveness correctly.

Minimum-validation-loss selection gives the same direction on all four headline configurations.
The result is not specific to the final checkpoint. Missing measurements remain explicit,
rather than being treated as zero detections.

The separately identified test-score Orthrus variant adapts its threshold to unlabelled test
scores. OCR-APT additionally consults attack identities during threshold fitting and enabled early
stopping, so final checkpoint selection does not make it a test-independent comparator.
Changing only thresholds leaves ranking metrics unchanged, but changing training need not.

\textbf{Full-campaign coverage at a finite budget.}
In a post-hoc comparison of four configurations on the three primary E3 hosts, Figure~\ref{fig:coverage-budget}
counts the members that reach every labelled campaign within $B$ ranked nodes. We display the two headline
word2vec configurations alongside Kairos and OCR-APT, whose THEIA contrast motivated this view, and retain
the same four configurations on every host. The reporting budget $B=100$ is one of the previously declared
budgets in Rule 2 (\S4), although selecting this comparison was post hoc, not a preplanned test.

On CLEARSCOPE, Kairos leads Orthrus on ADP (0.154 versus 0.120) but reaches full coverage within 100 nodes in 4 of 5 members, against 10 of 10.
Their ADP intervals overlap: this is an ordering of observed means, not a resolved performance difference.
The distinction also appears within a single model. One Orthrus member has ADP 0.501, yet its last campaign first appears at rank 6,754.
An alerting comparison must therefore state its coverage objective and budget, rather than treat ADP order as an operational preference.
OCR-APT's curves are a supervision diagnostic, not a fair comparison of unsupervised training procedures.

\subsection{Supporting case studies: capacity and padding}

Ranking also differs from alert delivery. We replay chronological two-hour score blocks with
validation-fitted thresholds, a member vote, a 24-hour node cooldown and a rolling alert cap.
These are policy replays conditional on scores being available at block end, not end-to-end
deployment measurements. On corrected THEIA E5, a ten-alert cap discards both labelled candidates
despite agreement from four of five members because strictly earlier blocks have already filled the queue.
Raising the cap recovers them. The two candidates describe one episode,
not independent attacks.

This is a counterexample, not a general failure rate. In the preregistered CADETS E5 test,
the primary ten-alert policy preserves both campaigns, while the five-alert sensitivity misses both. The supplementary padding test also distinguishes added workload from
evasion. Bilot et al. already observed that padding can increase alerts without evasion
\cite{bilot2025velox}. Our analysis separates original and injected endpoints: activity added to a
labelled process can itself trigger an alert, without demonstrating detection of its original payload.

\subsection{Encoder inputs and architectural claims}

Velox's study reports that word2vec, a linear encoder and edge-type prediction achieve the
highest mean ADP in its ablation on CADETS and THEIA (SC7). Its \S5.1 concludes that detection
"does not require modeling graph patterns"\cite{bilot2025velox}. A competitive linear encoder
establishes that message passing is unnecessary for that configuration's result, but it does not
establish that its inputs are graph-independent. We find a graph-derived signal in the released
reindexer, which assembles node features from an event batch. Our finding qualifies the attribution
to static attributes, not the observation that a non-GNN can remain competitive.

\textbf{Reindexer mechanism.} The default implementation averages source features into a buffer, then
adds destination features to that same buffer and divides by their occurrence count. If a node $v$ appears
in both roles, with a fixed feature vector $x_v$ and $d_v>0$ destination occurrences in the batch, the second
operation gives $\tilde x_v=(x_v+d_vx_v)/d_v=(1+1/d_v)x_v$. An encoder intended to read static attributes therefore
receives batch-degree information. The corrected operation averages concatenated source and destination
copies once, yielding $x_v$. Orthrus keeps the two representations separate, following TGN\cite{rossi2020tgn},
and is unaffected by this buffer-reuse defect. The paired retraining changes that operation under matched recorded settings.

This is not only a later PIDSMaker defect. Executing the original Velox release's loader,
reindexer and linear encoder at revision \texttt{ef9a9e2} maps constant features of 1.0 to 2.0 or 1.5 when
destination degree changes from one to two. The original tuple branch preserves 1.0.
This synthetic check establishes the released code path, not the original paper's training or effect sizes.
Our paired retraining resolves type-only decreases on both hosts and a word2vec decrease on CADETS,
while the THEIA word2vec effect remains unresolved (\S5.2). The correction removes an unintended encoder input,
not the detector's edge-type objective or incident-edge score aggregation.

The consequence is a narrower architectural conclusion, not a replacement winner. Corrected
hashed-path Velox remains competitive on CADETS (\S5.2), while the identical-input type-only controls
show that unintended degree scaling contributed to the released behavior. Together with the fixed-alert
and checkpoint contrasts, this separates three questions often compressed into one performance claim:
what was counted, how the reported model was chosen, and what information it consumed. Each requires
its own check even when the implementation is shared and the detector ordering is reproducible.

\section{STRICT: a measurement scheme for provenance-based intrusion detection}

STRICT turns the diagnosed mechanisms into six reporting requirements: Strict labels, Top-B budgets,
Replicated members, Independent selection, Chronological splits and Transparent scores. The principles are
established\cite{arp2022dosdonts,pendlebury2019tesseract,bouthillier2021variance}. What our measurements add is the
inference each requirement protects. Reproducing a score, or agreeing on which detector wins, does not establish
alarm-level precision, which depends on the label set and the scoring rule (Rule 1). It does not establish a
test-independent performance magnitude, which can be inflated by test-label selection (Rule 4). Nor does it establish
attribution to static attributes, which is confounded by batch structure leaking into the inputs (also Rule 4). The
controlled comparisons in \S3 and \S5 isolate these three inferences one at a time.

A shared implementation remains subject to the same checks, because common code can carry one selection or
scoring defect into every detector it runs. Rules 1, 2, 3 and 6 can be checked from released evaluation records,
whereas Rules 4 and 5 also need configuration and source evidence.

\textbf{Rule 1, Strict labels: exact matching against a declared target.} State the evaluation unit, the label
set and its version, and the labelled units active in the test period. An alert counts as a true positive only if the
alerted unit itself carries the target label. Such a match does not establish that the target is semantically correct. Report
neighbourhood-credit variants separately. Systems with different native outputs keep their own units, and a common
unit needs a specified mapping with corresponding labels.

\textbf{Rule 2, Top-B budgets: separate ranking from alert policy.} Report offline precision and campaign coverage
at predeclared budgets $B$, with a declared label-free tie-break, and report $B^{*}$, the number of alerts at which
every campaign is covered, together with tie-safe ADP. State the scored population and the observation period.
A top-$B$ cut limits retrospective selection but is not a chronological alert queue, and dividing $B$ by host-days
does not make it one. Report separately the native threshold and, where timestamps permit, a predeclared chronological
policy. This study uses $B=10,100,1000$, which are reporting points, not universal requirements.
Figure~\ref{fig:coverage-budget} shows why the summaries are not interchangeable: higher ADP can accompany a smaller fraction
of members that reach every campaign within the same budget.

Ties need their own report. A cut inside a block of equal scores, or the last block that $B^{*}$ must enter,
depends on the order within the block, which no score determines: on corrected type-only THEIA, mean $B^{*}$ ranges
from 23.9 to 46,639.3 without changing any score. Affected configurations are marked $^{\dagger}$, and
Table~\ref{tab:bstar-ties} gives their exact attainable bounds. Node-id order is label-free but not attack-neutral, whereas the
bounds cover every order, including randomized ones. Bootstrap intervals
answer a different question: they vary the sampled members, not the order within a tie.

\textbf{Rule 3, Replicated members: uncertainty from randomization.} Replicate training under a declared
randomization scheme, stating which sources vary and which stay fixed. Report the member count $n$, the member mean,
an uncertainty interval, and the minimum and maximum, with the interval method and its sampling unit. A best-of-$n$
result alone does not satisfy the rule. A single saved checkpoint cannot establish training variability, and
deterministic copies are not independent members. The rule concerns reporting variability, not reaching a universal
member count. We target at least five members and use bootstrap intervals\cite{efron1993bootstrap}. Descriptive ADP
summaries of fewer than ten members report the observed member range instead of an interval, while the five-pair paired
contrasts keep their bootstrap intervals as exploratory estimates. A member count
chosen for the precision of one mean does not certify a difference between systems, which needs uncertainty on the
difference itself.

\textbf{Rule 4, Independent selection: selection without test labels.} Declare the checkpoint, hyperparameter
and threshold procedures, and the data each may use, before test evaluation. Selection must not use test attack
labels. Held-out labelled validation is admissible when its supervision and its separation from test are explicit.
No particular checkpoint rule is required, and this study uses the final epoch and the minimum benign-validation loss.
Two further conditions are checked separately: metrics must use realizable, tie-safe operating points, and batching
or features must not expose structure that the model description excludes. Meeting one condition does not establish
the others. An undocumented selection counts as not established, neither a violation nor evidence of
test-independence.

\textbf{Rule 5, Chronological splits.} Training precedes validation and validation precedes test in wall-clock
time. When training uses days after an attack, the evaluation cannot establish prospective detection of that attack,
although the scores can still support retrospective ranking. Report each campaign's time to first alert under a stated streaming protocol.

\textbf{Rule 6, Transparent scores.} Release per-evaluation-unit, per-member scores with content hashes,
labels, configuration and the inputs needed to recompute claimed checkpoint or threshold decisions.
Scores and headline validation arrays are public on \href{https://huggingface.co/datasets/STRICT-APT/strict-score-record}{Hugging Face}.
\href{https://zenodo.org/records/22990291}{Zenodo} archives their hashes and member record, not score files.
\href{https://zenodo.org/records/22991158}{Code} is public, and anonymous score and code retrieval are verified. Rule 6 remains partial:
public threshold-selection inputs cover only the four headline configurations, and local verification
does not establish public access to its complete supporting record.

\textbf{What the requirements establish.} STRICT records whether the evidence supports a particular claim. It is
not an aggregate score. Test-label selection invalidates a claim of test-independent selection however many other
requirements hold, a nonchronological comparison can describe retrospective ranking but not prospective detection,
and a released score record permits checking the arithmetic but not independent training or label correctness. Each
requirement is reported as satisfied, violated, partial, not established or not applicable to the stated
claim, with its evidence, and an unreported procedure is not evidence of a violation. In Table~\ref{tab:checklist}, Rule 1
assesses exact matching in each system's declared native unit, so Kairos windows, CAPTAIN entities and ProvCTDG edges
can satisfy it. A common ranking across those units would still need the mapping that Rule 1 requires.

Applied to this study, the replicated reference comparisons satisfy Rules 1 to 3, while single-checkpoint
native reproductions do not establish training variability. Rule 4 is partial because the configured hyperparameters
and training lengths are inherited from PIDSMaker and their selection provenance is not established. The
configured CADETS split is nonchronological, and its chronological variant changes the campaign composition as well
as the chronology, so Rule 5 is also partial. Rule 6 is partial for the reason given above. STRICT compliance establishes
neither adversarial robustness nor deployment utility.

\section{STRICT evaluation}
\begin{table*}[!t]\centering\setlength{\tabcolsep}{4pt}\small
\begin{tabular}{lrrrr}
\toprule
Model & $n$ & ADP & P@100 & $B^{*}$ \\
\midrule
\addlinespace[2pt]\multicolumn{5}{l}{\textbf{CADETS E3}} \\
Flash $^{\dagger}$ & 5 & 0.20 [0.00, 0.34] & 0.02 & 40,826 \\
Kairos $^{\dagger}$ & 5 & 0.01 [0.00, 0.02] & 0.00 & 8,804 \\
MAGIC & 5 & 0.01 [0.00, 0.02] & 0.00 & 11,643 \\
NodLink & 5 & 0.32 [0.00, 0.67] & 0.01 & 2,331 \\
OCR-APT (label-dependent) & 5 & 0.69 [0.68, 0.69] & 0.04 & 13 \\
Orthrus (W2V, GNN) $^{\dagger}$ & 40 & 0.28 [0.22, 0.35] & 0.07 & 1,228 \\
Orthrus (test-score) & 5 & 0.33 [0.02, 0.63] & 0.08 & 545 \\
R-CAID $^{\ddagger}$ & 5 & 0.10 [0.01, 0.34] & 0.01 & 470 \\
ThreaTrace (port diagnostic) & 5 & 0.00 [0.00, 0.00] & 0.00 & 1,555 \\
Velox (W2V, linear) $^{\dagger}$ & 40 & 0.18 [0.10, 0.27] & 0.05 & 1,174 \\
\addlinespace[2pt]\multicolumn{5}{l}{\textbf{CLEARSCOPE E3}} \\
Flash $^{\dagger}$ & 5 & 0.00 [0.00, 0.00] & 0.00 & 11,311 \\
Kairos $^{\dagger}$ & 5 & 0.15 [0.01, 0.25] & 0.01 & 46 \\
MAGIC $^{\dagger}$ & 5 & 0.02 [0.01, 0.04] & 0.00 & 1,632 \\
NodLink & 5 & 0.01 [0.01, 0.02] & 0.00 & 2,519 \\
OCR-APT (label-dependent) & 5 & 0.00 [0.00, 0.00] & 0.00 & 20,854 \\
Orthrus (W2V, GNN) $^{\dagger}$ & 10 & 0.12 [0.05, 0.21] & 0.01 & 15 \\
Orthrus (test-score) $^{\dagger}$ & 5 & 0.00 [0.00, 0.01] & 0.00 & 352 \\
R-CAID $^{\ddagger}$ $^{\dagger}$ & 5 & 0.00 [0.00, 0.00] & 0.00 & 2,328 \\
ThreaTrace (port diagnostic) $^{\dagger}$ & 5 & 0.00 [0.00, 0.01] & 0.00 & 15,806 \\
Velox (W2V, linear) $^{\dagger}$ & 5 & 0.05 [0.01, 0.12] & 0.01 & 47 \\
\addlinespace[2pt]\multicolumn{5}{l}{\textbf{THEIA E3}} \\
Flash $^{\dagger}$ & 5 & 0.00 [0.00, 0.01] & 0.01 & 3,003 \\
Kairos $^{\dagger}$ & 5 & 0.23 [0.13, 0.26] & 0.01 & 1,600 \\
MAGIC & 5 & 0.00 [0.00, 0.00] & 0.00 & 197,050 \\
NodLink & 5 & 0.00 [0.00, 0.00] & 0.00 & 149,215 \\
OCR-APT (label-dependent) & 5 & 0.12 [0.12, 0.12] & 0.04 & 52 \\
Orthrus (W2V, GNN) $^{\dagger}$ & 40 & 0.27 [0.21, 0.34] & 0.03 & 15,377 \\
Orthrus (test-score) $^{\dagger}$ & 5 & 0.10 [0.02, 0.22] & 0.03 & 175 \\
R-CAID $^{\ddagger}$ & 5 & 0.00 [0.00, 0.01] & 0.00 & 1,570 \\
ThreaTrace (port diagnostic) & 5 & 0.00 [0.00, 0.00] & 0.00 & 201,903 \\
Velox (W2V, linear) $^{\dagger}$ & 40 & 0.42 [0.32, 0.53] & 0.10 & 443 \\
\bottomrule
\end{tabular}
\caption{Descriptive STRICT reference comparison: final checkpoints, strict Orthrus targets, grouped by E3 host. ADP: mean [95 \% member-bootstrap interval for $n\geq10$, observed member range otherwise]. P@100 and $B^{*}$: means with node-id tie-breaking. $^{\dagger}$ marks tie-sensitive coverage cost. Higher ADP/P@100 and lower $B^{*}$ are preferred. W2V means word2vec. Velox is corrected. OCR-APT is label-dependent. ThreaTrace's port reverses native confidence polarity. $^{\ddagger}$ R-CAID scores a different population on each host and is not rank-comparable. Repeated outcomes are not independent confirmations. 0.00 need not mean zero.}\label{tab:strict-results}
\end{table*}

This section applies \S4's requirements to a descriptive fixed-host comparison with test-independent checkpoint selection,
fixed-output target sensitivity, and seed-matched retraining comparisons of the reindexer correction.

Table~\ref{tab:strict-results} compares nine detector families on CADETS E3, CLEARSCOPE E3 and THEIA E3,
the hosts with the broadest system coverage, with the Orthrus test-score variant identified separately.
The frozen comparison roster uses word2vec for both Velox and Orthrus, holding their feature method fixed
instead of selecting each family's best tested features. Corrected hashed-path Velox, reported in \S5.2,
reaches ADP 1.00 on CADETS. The table therefore compares configurations, not each family's attainable performance.
All rows use PIDSMaker's implementations and the final checkpoint, with native results reported separately.
OCR-APT is a supervision diagnostic. R-CAID's scored populations differ from the other
systems' (\S3.4), so its rows are reported separately from any comparison of detector ordering.
Final checkpoint selection removes test-label checkpoint choice but does not equalize training
schedules or establish the provenance of inherited tuning. The table describes these
configurations, not a significant ranking of detector families.

ThreaTrace is also an implementation diagnostic: the port flags a confident, correct node-type
prediction, whereas the native implementation removes it as benign. A source-level test reproduces
this polarity difference, and all fifteen final-checkpoint E3 executions in the port alert on every scored
node. These rows characterize the inspected port, not native ThreaTrace's detection quality.

Score replay reproduces every measured main-table row. Checkpoint and threshold
reconstruction additionally covers the four headline word2vec configurations. This verifies
numerical replay, not raw-data retraining or semantic label validity.

\textbf{Additional E5 hosts.} With word2vec features and five members per configuration,
corrected Velox has higher final-checkpoint mean ADP than Orthrus on THEIA E5 (0.275 versus 0.003)
and CLEARSCOPE E5 (0.288 versus 0.236). THEIA Velox varies substantially across members.
These results extend the descriptive comparison beyond E3 without establishing a general
advantage for the linear encoder, which trails Orthrus on CADETS E3.

\subsection{Target sensitivity and label validity}

\textbf{Do the contrasts survive another target?} We recompute twenty checkpoint and reindexer
contrasts under both node-label lists, holding scores, alarms and populations fixed.
The final-checkpoint CADETS word2vec correction lowers P@100 by 0.0595 under Orthrus targets and 0.0635 under
represented ThreaTrace targets. Because other effects depend on the target and budget, no campaign-ADP robustness is
claimed. Most ThreaTrace CADETS targets lie outside the scored population, whereas almost all
THEIA targets are represented. These are not
evaluations on the complete native target populations.

\subsection{The corrected controls}
\begin{table*}[t]\centering\setlength{\tabcolsep}{4pt}\small
\textbf{(a) Historical recorded-seed comparisons}\par\smallskip
\begin{tabular}{llr}
\toprule
Host / features & Pairs & Final $\Delta$ADP \\
\midrule
CADETS / Type only & 10 & -0.385 [-0.532, -0.235] \\
CADETS / word2vec & 40 & -0.376 [-0.472, -0.269] \\
CADETS / Hashed path & 10 & +0.015 [+0.007, +0.024] \\
THEIA / Type only & 10 & -0.453 [-0.526, -0.385] \\
THEIA / word2vec & 40 & +0.006 [-0.112, +0.115] \\
THEIA / Hashed path & 10 & +0.065 [-0.142, +0.264] \\
\bottomrule
\end{tabular}
\par\medskip\textbf{(b) Identical-input type-only controls}\par\smallskip
\begin{tabular}{lrrrr}
\toprule
Host & Released ADP & Corrected ADP & $\Delta$ADP [range] & Decreased pairs \\
\midrule
CADETS & 0.220 & 0.0013 & -0.219 [-0.475, -0.042] & 5/5 \\
THEIA & 0.647 & 0.0005 & -0.647 [-0.749, -0.417] & 5/5 \\
\bottomrule
\end{tabular}
\caption{Reindexer correction at the final checkpoint. $\Delta$ADP is corrected minus released. (a) Historical members paired by recorded seeds and settings: mean [95 \% bootstrap interval over pairs], without established input identity. (b) Five identical-input type-only pairs per host: condition means and paired mean [observed minimum, maximum], not confidence intervals. Inputs, initial parameters, optimizer state and random states match within each pair, and actual and configured dropout are 0.3. Correction lowers ADP in every pair. THEIA has five trainings but three distinct released ADP outcomes. E3 hosts, with a nonchronological CADETS split. Designs are not pooled.}\label{tab:reindexer-core}
\end{table*}

\begin{table*}[t]\centering\setlength{\tabcolsep}{4pt}\small
\begin{tabular}{>{\raggedright\arraybackslash}p{0.15\textwidth}>{\raggedright\arraybackslash}p{0.32\textwidth}>{\raggedright\arraybackslash}p{0.44\textwidth}}
\toprule
Prior work & Established contribution & Inference examined here \\
\midrule
Bilot et al.\cite{bilot2025velox} & Evaluation pitfalls and a competitive linear encoder. & Does non-GNN competitiveness isolate static-attribute sufficiency? Released preprocessing injects degree information, and identical-input controls isolate its effect (\S3.6, \S5.2). \\
PIDSMaker\cite{bilot2026pidsmaker} & Common implementations and repeated-experiment support. & Does a reproduced ordering establish test-independent performance? Fixed-member reselection changes magnitude even when the headline order survives (\S3.4). \\
Guerra et al.\cite{guerra2026benchmarks} & Audited benchmarks, shared targets, validation-only selection and distinct alerting/investigation metrics. & Do shared targets make reported precisions comparable? Fixed-alert comparisons separate supplied labels from scoring credit and expose their interaction (\S3). \\
\bottomrule
\end{tabular}
\caption{Closest studies and the additional controlled evidence. Rows identify the comparison relevant to each study, not an exhaustive inventory or a claim of exclusive scope.}\label{tab:closest-work}
\end{table*}

\textbf{Corrected controls.} The main comparisons use corrected Velox, while default-reindexer rows remain
diagnostics of the unintended degree signal (\S3). The word2vec controls on CADETS and THEIA use forty
members each. We report final-checkpoint scores and evaluate alerts with each member's native
maximum-validation threshold. Settings match apart from the reindexer,
but the two conditions train different models.

\textbf{Pairing by seed.} A corrected configuration and its default twin are launched with identical settings apart
from the reindexer, so members with equal featurization and training seeds form pairs.
On CADETS, correction lowers word2vec ADP under both checkpoint rules, but on THEIA the effect remains
unresolved. Table~\ref{tab:reindexer-core} gives the final-checkpoint effects.
Matching recorded seeds and settings does not prove identical feature realizations or deterministic
execution. Unlike checkpoint reselection, these comparisons train different models.

\textbf{Direction depends on host and checkpoint.} Smaller word2vec extensions show an increase
on CLEARSCOPE and a checkpoint-dependent result on OpTC-201. Other OpTC and E5 effects remain
unresolved. The defect changes the encoder's information, but correcting it
does not uniformly reduce ranking quality.

\textbf{Identical-input controls on two hosts.} Five initialization pairs per host hold complete
pre-reindexer inputs, initial parameters, optimizer state and random states fixed within each pair.
Actual and configured dropout agree. Correction lowers type-only ADP in every pair on both hosts.
THEIA's five trainings yield three distinct released ADP outcomes, so repeated values
do not provide five distinct confirmations.
The identical-input block in Table~\ref{tab:reindexer-core} gives the mean and range of paired effects.
Corrected ADP stays near zero on both hosts. These controls isolate the operation on retained
inputs, not historical word2vec effect sizes. The decrease measures the contribution of unintended
input scaling, showing why correcting an implementation need not improve its detection performance.
An earlier three-pair experiment with a dropout deviation remains separate.

\textbf{Type-only and named features.} Historical type-only correction removes nearly all
final-checkpoint ADP on both hosts. The CADETS effect is smaller under validation-loss selection because
that rule often selects an early, already weak checkpoint. Corrected hashed-path features retain
strong CADETS ranking. Their use of names, endpoint pairs and incident-edge losses precludes
attributing this result to static attributes alone. Table~\ref{tab:reindexer-core} separates feature conditions
and control designs rather than presenting one architecture ordering.

\section{Limitations}

Our comparisons cover DARPA E3/E5, OpTC, ATLASv2 and Carbanak v2, but not every detector on every
dataset. Source findings concern the inspected revisions, and replication intervals describe the declared
randomization on these hosts, not variability across organizations or future attacks. Native evaluation units
remain separate where no validated mapping exists.

Exact matching makes scoring reproducible, not labels authoritative. The selected 39-case
language-model review supports interpretations of those alerts, not population precision or independent
ground truth. Human agreement is unmeasured, and mechanical response validity is not semantic correctness.

Offline ranking and chronological alerting answer different questions. The capacity replays use
saved scores assumed available at block end and do not establish prospective end-to-end detection with
causal feature fitting and graph state. Alert counts measure workload, not analyst time or investigation
quality, and fixed-budget reporting does not resolve threshold transfer (\S3.5).

\section{Related work}

\textbf{Detection and investigation.} Provenance analysis supports several distinct security tasks
\cite{inam2023sok}. Whole-graph methods such as StreamSpot\cite{manzoor2016streamspot} and Unicorn
\cite{han2020unicorn} identify anomalous graph behavior, whereas node- and window-level detectors localize
different units of activity. Systems such as SLEUTH, HOLMES, POIROT and ATLAS use provenance for
investigation, attack correlation or reconstruction\cite{hossain2017sleuth,milajerdi2019holmes,milajerdi2019poirot,alsaheel2021atlas}.
Their outputs need not answer the same question. Finding a
useful path to an attack can aid investigation without making every alert on that path malicious.
Conversely, exact-node scoring can miss the value of contextual evidence. Our fixed-alert analysis
separates these interpretations: it tests the meaning of reported precision, not whether neighbourhood
context is useful to an analyst.

\textbf{Closest evaluation studies.} Bilot et al., PIDSMaker and Guerra et al. establish much of the
evaluation practice summarized by STRICT (Table~\ref{tab:closest-work}). Our contribution is not the principles
but controlled evidence for three inferences that a reproduced score or an agreed winner leaves open:
alarm-level precision, test-independent performance magnitude and attribution to static attributes.
The distinction concerns what an experiment establishes, rather than how many systems it compares.
A common evaluator makes results easier to compare, but its counting, selection and input-construction
operations also require validation.

Bilot et al. identify shortcomings in provenance-detector evaluation, including inconsistent
granularity, data snooping and instability, and show that the simpler Velox encoder can be competitive
\cite{bilot2025velox}. We build on that critique rather than treating architectural complexity as evidence
of superiority. Our reindexer diagnosis asks a narrower question: did the released linear encoder
consume only the intended static attributes? Its input construction introduces degree-dependent
scaling. Identical-input controls isolate that operation, while historical feature comparisons show
that its effect depends on the feature representation and host. This qualifies attribution to static
attributes without refuting non-GNN competitiveness. An implementation correction and a performance
improvement are different claims, since removing an unintended signal can lower measured detection quality.

PIDSMaker standardizes preprocessing and labels, provides reusable detector components, and
supports tuning, ablations and repeated-experiment stability analysis\cite{bilot2026pidsmaker}. These
capabilities make controlled comparisons practical and underpin much of our study. Rather than propose
another common implementation, we examine whether its measurements support the inference
attached to them. Reselecting checkpoints from the same trained members isolates selection from
training variation, while reconstructing counts from unchanged alarms isolates the evaluator from the
detector. Native-release checks then distinguish findings in the common implementation from behavior
in upstream artifacts. Agreement between implementations is informative, but a shared operation can
also reproduce the same measurement error.

\textbf{Concurrent benchmark analysis.} Guerra et al.'s concurrent preprint precedes ours. It audits
benchmark quality, relates semantic signal quality to detector performance, and finds a lexical
allowlist competitive at key operating points\cite{guerra2026benchmarks}. Its comparison uses shared
targets, separates alerting from investigation, and selects primary E3 checkpoints using labelled
validation attacks. Our primary checkpoint is fixed, with benign-validation loss as a sensitivity
analysis. This supervision difference is not our novelty claim. Both studies caution against reading
benchmark success as evidence of general attack understanding, but they examine different mechanisms.
Whereas benchmark quality can determine which signals are learnable, our controls test how counting rules,
model selection and unintended encoder inputs affect the interpretation of a measured result.

Shared targets alone do not settle that interpretation. Holding alarms fixed while crossing
target labels with the scoring rule reveals an interaction: changing the target set need not change
precision in the same direction under exact matching and neighbourhood credit. The result does not
establish that one supplied label set is semantically authoritative. It shows why target harmonization
and scorer validation are separate requirements. Similarly, an unchanged ordering of headline
detectors does not establish that their reported performance magnitudes survive test-independent
checkpoint selection. These controlled contrasts complement benchmark audits rather than replacing
their assessment of the underlying data.

\textbf{Selection bias and experimental variation.} Security-ML guidance already treats leakage,
sampling bias and unsuitable metrics as threats to validity\cite{arp2022dosdonts}. TESSERACT demonstrates
the importance of temporal and distributional constraints in malware evaluation
\cite{pendlebury2019tesseract}. More generally, adaptive reuse of a holdout can overfit the evaluation
sample\cite{dwork2015holdout}. STRICT applies these established concerns to the concrete decisions that
produce a provenance-detector result. A test-independent checkpoint is one such decision, not a
certificate for the whole experiment: feature fitting, graph construction, split chronology and
threshold calibration have their own information boundaries. This is why we distinguish
test-independent checkpoint selection from a fully prospective detector deployment.

Repeated training also addresses a different uncertainty from benchmark generalization.
Bouthillier et al. show that randomness in data sampling, initialization and model selection affects
benchmark comparisons\cite{bouthillier2021variance}. Our member-level intervals describe the declared
randomization on fixed hosts, not variation across future organizations. The strength of a paired comparison
also depends on what is held fixed. Matching recorded seeds does not prove that two conditions consumed
identical realized features, whereas checkpoint reselection leaves the trained member unchanged.
The identical-input reindexer experiments additionally fix prepared inputs and initial states within
each pair. Keeping these designs separate makes the source of a measured difference explicit.

\textbf{Calibration and distribution shift.} Transcend identifies aging malware classifiers through
statistical comparison with their training data\cite{jordaney2017transcend}, and TRANSCENDENT develops
conformal rejection for classification under drift\cite{barbero2022transcendent}. These methods make
distributional assumptions and rejection decisions explicit. Our calibration analysis asks why a
particular benign-only rule can still generate unstable alert workloads. Taking the largest validation
score makes the effective tail probability depend on the validation sample size, even under
exchangeability. Changes in the population or observation period add separate concerns. The fixed-model
subsampling experiment exposes the sample-size dependence without retraining or changing test scores,
while the composition analysis examines a different member's false alerts. Neither observation establishes
that benign-only calibration is intrinsically ineffective or that a drift detector would resolve the
measured failure.

\textbf{Workload and operational interpretation.} NoDoze prioritizes alerts using their provenance
context\cite{hassan2019nodoze}, and Dong et al.'s industrial study identifies both alarm-triage and
interpretation costs as obstacles to adoption\cite{dong2023arewethereyet}. Earlier network-intrusion
detection work likewise cautions against equating benchmark classification with operational utility
\cite{sommer2010outside}. Our fixed-budget coverage and chronological capacity comparisons address a
limited part of this problem: whether labelled campaign evidence remains accessible under a stated
review budget. They do not measure analyst time or the quality of an investigation. A higher average
ranking score may coexist with an uncovered campaign, while a daily capacity limit can suppress a
candidate that exists in the offline ranking. Reporting these outcomes separately avoids interpreting
one metric as a substitute for the entire response workflow.

\textbf{Adversarial robustness.} Goyal et al. study mimicry against provenance anomaly detectors by
adding benign-looking behavior while preserving the underlying attack\cite{goyal2023mimicry}. Bilot et al.
also examine padding and observe cases in which it increases alerts rather than achieving evasion
\cite{bilot2025velox}. Our supporting padding experiment concerns the attribution of those alerts.
An injected event can touch an already labelled process, so renewed coverage of that process need not
mean that the original malicious activity was detected. Separating original and injected endpoints
makes this ambiguity visible. This complements evasion testing: changing the attack context and
changing what receives detection credit must not be conflated. The offline experiment does not
establish a live-system evasion capability.

Reproducibility studies examine whether published provenance-detector results can be
recovered from released code and experimental descriptions\cite{abrar2025reproducibility}. Our audit
addresses the next question as well: what does a recovered result mean? Retaining score vectors,
target mappings, selection decisions and counted alarms allows numerical reconstruction to be
separated from semantic label validation and end-to-end training reproduction. STRICT records these
boundaries alongside the measurements. Reproducibility is essential to inspect a claim, but does not
by itself establish that the claim follows from the reproduced number.

\section{Conclusion}

Reliable provenance-detector evaluation requires more than reproducing a score or agreeing on
the winning model. Our audit traces three distinct sources of interpretation error in released
systems. Fixed-alert comparisons separate supplied targets from the credit assigned by the scorer.
Checkpoint reselection measures how test-label access changes reported performance without retraining
the detector. Reindexer controls establish that an ostensibly static-attribute encoder receives an
unintended degree-dependent signal. These experiments identify which inference changes when the
relevant operation changes, rather than attributing every discrepancy to the detector architecture.

The central lesson is that stable detector ordering can coexist with unstable conclusions about
performance magnitude and mechanism. Non-GNN competitiveness does not establish that static attributes
alone explain the result, and high ranking quality does not ensure that every campaign is represented
within a fixed review budget. The corrected controls reinforce this distinction: removing unintended
information can lower ADP without making the correction invalid. Evaluation must distinguish the
quality of a detector's decisions from the validity of the explanation offered for them.

STRICT connects these findings to checkable reporting requirements: identify the scored unit
and targets, preserve test-independent selection, make ties and randomization explicit, report alerts
against a stated budget, and retain the evidence needed to reconstruct each measurement. The resulting
comparisons remain conditional on the benchmark targets and inspected implementations. They provide
a more precise basis for comparison, not a universal detector ranking or evidence of deployment readiness.
Before crediting an architecture with a benchmark result, establish what was counted,
how the reported model was chosen, and what information it actually consumed.

\flushcolsend\clearpage\label{bib:start}
\bibliographystyle{plain}
\bibliography{refs}
\section*{Ethical Considerations}

This study analyzes pre-existing DARPA Transparent Computing, OpTC, ATLASv2 and Carbanak v2 benchmark logs and released research
software. It does not deploy a detector on a live organization, collect new user telemetry, or intervene in an
operational security workflow. Stakeholders nevertheless include dataset stewards, detector authors and artifact
maintainers, security analysts who may rely on reported performance, and researchers who may build on the audited evaluation path.

Findings about released implementations could be misread as claims about their authors' intent.
We report observable behavior at pinned revisions, quote decisive source lines, distinguish direct
test-label contact from undocumented constants, and separate common implementations from native artifacts.
We retain negative and contradictory results and limit source findings to the inspected revisions.

The released material evaluates research-benchmark attacks and does not provide a new path to compromise a
live system. Detailed failure analyses could still help an attacker reason about anomaly-detector blind spots, but
the countervailing benefit is that defenders and reviewers can detect invalid evaluation claims. We minimize the
risk by restricting our tests to offline evaluation rather than weaponized host exploits, and by following the source
datasets' redistribution and licensing terms instead of republishing restricted raw logs.

The padding generator is dual-use: synthetic benign-looking activity could help study alert flooding
or suppression by a bounded queue. Our tests modify offline benchmark graphs, require campaign knowledge,
and neither execute payloads on hosts nor optimize against live detector feedback. We report the fixed
experimental budgets, seeds and endpoint separation needed to reproduce this measurement failure,
without presenting the generator as a deployable evasion tool. These limits reduce, but do not eliminate, misuse risk.

Before either initial rater ran, we amended the preregistered two-human review to use language models (Appendix A).
The frozen 39-case packet was reviewed by three language-model raters from two families. An exploratory six-family extension subsequently collected 234 responses and made one additional request for each of the 34 invalid ones, retaining invalid responses and reporting coverage only. Neither assessment involved human participants or changed the dataset labels. The study record retains the amendment, prompts, transcript tool-use audit and hashed forms, although retention does not imply public access to these materials. The judgments are research measurements, not operational incident-response decisions.

\textbf{External model processing.} The reviewed cases derive from the released DARPA THEIA E5
benchmark, not live organizational telemetry. Requests contain the rubric and selected case events with
opaque case identifiers and relative times, excluding detector scores, labels, sampling roles and prior
judgments. Because original command, path and network-address strings remain necessary context, this is not
full anonymization. The six-family collection used author-approved external services, with requests
restricted to endpoints advertised as zero-retention, with data collection and tools disabled.
Approved fallback routes stayed within explicit same-model provider allowlists.
These are recorded request controls, not an independent audit of provider
retention. Credentials and unrelated project material were not included in model-visible evidence.

\section*{Open Science}

Evaluation and analysis code is available under Apache 2.0 on
\href{https://zenodo.org/records/22991158}{Zenodo}. Released scores and headline validation arrays are
available under CC BY 4.0 on \href{https://huggingface.co/datasets/STRICT-APT/strict-score-record}{Hugging Face}.
A separate \href{https://zenodo.org/records/22990291}{Zenodo record} archives metadata and validation arrays,
not the score files. Each release identifies its included configurations and score hashes.

The public material supports recomputing metrics from released scores and reconstructing thresholds
for the four headline configurations. Other validation losses, raw-derived training inputs
and full execution records are private, so Rule 6 remains partial. Raw benchmark data and restricted
third-party implementations are not redistributed. Our local numerical checks cover more evidence
than this public subset and do not establish independent raw-data retraining.

Preregistration refers to repository plans fixed before the specified experiments' outcomes,
not registration in a public registry. These plans and their amendments are retained privately,
and the whole study was not preregistered.

\appendix
\raggedend
\section{Labels and native evaluations}
\textbf{Precision aggregation.} Let $A_i=\mathrm{TP}_i+\mathrm{FP}_i$ be the scorer's denominator and let $I$ contain the $m$ members with $A_i>0$, out of $n$ members. Conditional mean precision is $\bar P_{+}=m^{-1}\sum_{i\in I}\mathrm{TP}_i/A_i$. The zero-filled mean is $\bar P_{0}=n^{-1}\sum_{i\in I}\mathrm{TP}_i/A_i=(m/n)\bar P_{+}$ when $m>0$. Pooled precision is $P_{\mathrm{pool}}=\sum_i \mathrm{TP}_i/\sum_i A_i$: it weights members by $A_i$, whereas $\bar P_{+}$ weights members with defined precision equally. Under strict scoring $A_i$ counts alerts, but neighbourhood-credit scoring changes these counts (\S3). When $A_i=0$, precision is undefined. Assigning it zero in $\bar P_{0}$ is a reporting convention, not an observed false alert. If $m=0$, $\bar P_{+}$ is undefined and we report zero for $\bar P_{0}$ and, by convention, the pooled table entry. TP and FP means always include all $n$ members.

\textbf{How ADP and the 2-hop relaxation are computed.} ADP walks the scores from high to low. For each distinct precision the procedure keeps the largest fraction of campaigns detected. It integrates these points by the trapezoidal rule in increasing precision, including $(0,0)$ and stopping at the largest attained precision without extrapolating to 1. Campaign coverage is a fraction in $[0,1]$. Tie-safety means the walk may stop only at thresholds a score can realize, so all nodes sharing a score enter together. The released implementation instead walks tied nodes one at a time. The 2-hop relaxation counts a missed labelled node within two hops of a detected one as detected, and drops a false positive within two hops of any labelled node.

\textbf{Native identity checks.} Replaying MAGIC's parser reproduces the retained CADETS graph's node order, edges and type partitions, and exported scores match the native cache row for row. Its 12,846 malicious indices map into the ThreaTrace UUID list. Six further in-population labels are unnamed pipes excluded by the parser. These checks establish mapping consistency, not fresh inference or label semantics.

The released ThreaTrace list contains 12,858 node UUIDs, of which 12,852 resolve in its test shard. The list does not supply timestamps, campaign membership or an adjudication history, so we cannot establish whether its breadth comes from malicious actions, attack-related context or graph expansion. Neighbourhood credit is a separate executable scorer operation, not evidence about how the list was constructed. Our fixed-alert comparison changes the supplied target without assigning a semantic explanation to this label-set difference.

\textbf{Label prevalence and lift.} Precision depends on target prevalence as well as alert concentration \cite{axelsson2000baserate}. In the native CADETS population, the labelled fractions are $\pi_T=0.036$ and $\pi_O=0.00013$. Lift is precision divided by $\pi_L$, the expected precision of a random alert. MAGIC's lift falls from 26 to 4.6: its 1,600-fold precision drop combines a 280-fold base-rate change and a 5.7-fold loss of concentration. ThreaTrace's strict lift instead rises from 0.20 to 70. Neither result adjudicates the target lists.

\textbf{Different evaluated populations.} The native CADETS comparison scores 357,173 nodes over the released test shard. The separate PIDSMaker comparison scores 281,585 nodes and restricts ThreaTrace targets to nodes active inside the configured attack windows: 40 targets, versus 12,852 in the native ThreaTrace comparison. Orthrus contributes 68 versus 46 targets in these respective populations. Each controlled contrast holds its own population fixed, so effects cannot be combined across the two. Table~\ref{tab:official} retains each native artifact's own comparison, separately from the PIDSMaker results.

\textbf{Matched adjacency analysis.} An alert is adjacent if a directed, time-respecting path of at most two edges connects it to a labelled node within the observed event order. Each false positive has twenty benign non-alert controls matched on node type and logarithmic degree bin. Across forty corrected members, mean adjacent-FP/control fractions are 0.79/0.82 on CADETS and 0.85/0.71 on THEIA. These are descriptive means, not paired effect intervals. Controls exclude every node alerted by any member, so their population depends on ensemble membership. This comparison establishes neither semantic usefulness nor equivalence to the null.

\textbf{Fixed-output target sensitivity.} Table~\ref{tab:target-sensitivity} gives all twenty declared P@100 contrasts: checkpoint changes in the four headline configurations and six reindexer comparisons under both label-free rules. Scores, populations and original Orthrus-based checkpoint choices stay fixed. The alternative-target results therefore do not estimate selection inflation for a ThreaTrace-specific selector. The experimental record also retains P@10, P@1000 and labelled fractions of saved alarms.

Every scored node has an unambiguous retained UUID, and joining the configured Orthrus lists exactly reconstructs the saved labels. CADETS has 281,585 scored nodes with 68 Orthrus targets. Among 12,858 unique ThreaTrace targets, 40 occur in this population, 12,811 map outside it and seven are unmapped. THEIA has 700,902 scored nodes with 118 Orthrus targets. Its ThreaTrace list contains 25,363 entries and 25,358 unique targets, of which 25,342 are scored, twelve map outside and four are unmapped. These coverage differences prevent interpreting cross-target precision as a detector-quality comparison.

Intervals resample paired members, not hosts. Separate target-specific intervals do not test an interaction or equivalence between label sets. P@B uses node-id tie-breaking, while the retained exact individual tie bounds and contrast envelopes allow independent tie orders, not one shared global order. Saved-alarm fractions include only pairs where both members alert, with that count reported beside all pairs. Precision remains undefined for silence. The alternative list has no campaign partition, so neither campaign ADP nor full-coverage cost is recomputed. This analysis establishes neither label truth nor deployment utility.

\textbf{Language-model case review.} The frozen 39-case packet contains fifteen historical capacity cases, twelve stratified temporal-negative alerts and twelve matched benign controls from the default-reindexer THEIA E5 stream. Raters receive a three-class rubric and bounded event excerpts with opaque identifiers, without detector scores, labels or sampling groups, although event names can still reveal the benchmark. An amendment before either initial review substitutes models for the proposed human review. No human agreement is measured.

The two initial raters agree on 38 of 39 cases, with Cohen's $\kappa=0.93$ when uncertain is a class. A model from another family agrees on the fifteen capacity cases: two candidates judged malicious and thirteen predecessors benign, all concerning one browser/file episode. One predecessor explanation contains an event-citation error, and a suspected label gap remains disputed. No judgment changes the benchmark targets. Cross-family agreement on these restricted inputs does not establish semantic truth.

\textbf{Exploratory panel.} All 234 planned responses returned, and 200 met the amended mechanical requirements and 198 the original requirements. These checks assess response structure and event citations, not interpretation accuracy. The predeclared constructed-case criterion failed, leading to an exploratory amendment. The all-responses-valid criterion also failed, so no full-panel agreement or aggregate labels are reported.

An additional assessment of each of the 34 initially invalid responses returns 33 replies: thirteen satisfy amended requirements, ten of them the original requirements, while twenty remain invalid and one request returns no judgment. No later reply replaces an initial one. This selected failure diagnostic does not establish repeatability.

The case-control sample cannot estimate prevalence, recall or operational precision. Correlated errors, benchmark recognition and limited context remain possible. The supported capacity result is suppression of labelled candidates, not expert validation of missed malicious activity.
\begin{table*}[tp]\centering\setlength{\tabcolsep}{2pt}\footnotesize
\begin{tabular}{lllllll}
\toprule
system & host & model & alerts & own metric P / R & strict Orthrus TP / FP / P & strict P@100 \\
\midrule
Flash & CADETS & released weights & 19,774 & 0.928 / 1.000 & 20 / 19,754 / 0.0010 & --- \\
Flash & CADETS & member 0 & 22,428 & 0.816 / 1.000 & 18 / 22,410 / 0.0008 & --- \\
Flash & CADETS & member 1 & 20,587 & 0.889 / 1.000 & 17 / 20,570 / 0.0008 & --- \\
Flash & CADETS & member 2 & 5,271 & 0.987 / 1.000 & 8 / 5,263 / 0.0015 & --- \\
Flash & CADETS & member 3 & 22,640 & 0.822 / 1.000 & 17 / 22,623 / 0.0008 & --- \\
Flash & CADETS & member 4 & 10,789 & 0.822 / 1.000 & 17 / 10,772 / 0.0016 & --- \\
Flash & THEIA & released weights & 24,996 & 0.989 / 0.999 & 3 / 24,993 / 0.0001 & --- \\
Flash & THEIA & member 0 & 28,023 & 0.739 / 0.999 & 2 / 28,021 / 0.0001 & --- \\
Flash & THEIA & member 1 & 22,317 & 0.949 / 0.999 & 3 / 22,314 / 0.0001 & --- \\
Flash & THEIA & member 2 & 24,265 & 0.946 / 0.999 & 1 / 24,264 / 0.0000 & --- \\
Flash & THEIA & member 3 & 32,014 & 0.740 / 0.999 & 4 / 32,010 / 0.0001 & --- \\
Flash & THEIA & member 4 & 23,312 & 0.963 / 0.999 & 0 / 23,312 / 0.0000 & --- \\
MAGIC & CADETS & released checkpoint & 13,575 & 0.944 / 0.998 & 8 / 13,567 / 0.0006 & 0.05 \\
MAGIC & CADETS & member 0 & 13,637 & 0.940 / 0.998 & 9 / 13,628 / 0.0007 & 0.05 \\
MAGIC & CADETS & member 1 & 18,342 & 0.699 / 0.998 & 21 / 18,321 / 0.0011 & 0.05 \\
MAGIC & CADETS & member 2 & 14,333 & 0.894 / 0.998 & 12 / 14,321 / 0.0008 & 0.05 \\
MAGIC & CADETS & member 3 & 13,763 & 0.931 / 0.998 & 8 / 13,755 / 0.0006 & 0.05 \\
MAGIC & CADETS & member 4 & 15,795 & 0.811 / 0.998 & 10 / 15,785 / 0.0006 & 0.05 \\
MAGIC & THEIA & released checkpoint & 25,774 & 0.982 / 1.000 & 60 / 25,714 / 0.0023 & 0.10 \\
CAPTAIN & CADETS & default parameters & 3,900 & 0.328 / 0.846 & 40 / 3,860 / 0.0103 & --- \\
CAPTAIN & CADETS & trained parameters & 264 & 0.440 / 0.846 & 39 / 225 / 0.1477 & --- \\
ThreaTrace & CADETS & released example models & 994 & 0.938 / 1.000 & 9 / 985 / 0.0091 & --- \\
ThreaTrace & CADETS & member 0 & 3,331 & 0.819 / 1.000 & 11 / 3,320 / 0.0033 & --- \\
ThreaTrace & CADETS & member 1 & 2,724 & 0.838 / 1.000 & 4 / 2,720 / 0.0015 & --- \\
ThreaTrace & CADETS & member 2 & 9,698 & 0.705 / 1.000 & 27 / 9,671 / 0.0028 & --- \\
ThreaTrace & CADETS & member 3 & 4,660 & 0.755 / 0.999 & 5 / 4,655 / 0.0011 & --- \\
ThreaTrace & CADETS & member 4 & 8,632 & 0.701 / 1.000 & 25 / 8,607 / 0.0029 & --- \\
\bottomrule
\end{tabular}
\caption{Released detectors executed outside PIDSMaker with their own code on the raw CDM shards. Native precision beside strict Orthrus-node precision on the same alerts. MAGIC changes labels only, while Flash and ThreaTrace change both labels and scorer. CAPTAIN retains its native name-level metric. The fixed-alert factorial isolates the scorer effect separately. Released and retrained members remain distinct. CAPTAIN's two settings are not training repetitions.}\label{tab:official}
\end{table*}

\begin{table*}[tp]\centering\setlength{\tabcolsep}{2pt}\small
\begin{tabular}{llrrr}
\toprule
Change & Checkpoint & Pairs & Orthrus $\Delta$P@100 & ThreaTrace $\Delta$P@100 \\
\midrule
\addlinespace[2pt]\multicolumn{5}{l}{\textbf{CADETS E3}} \\
Checkpoint / Orthrus & Final & 40 & -0.031 [-0.046, -0.015] & -0.020 [-0.031, -0.009] \\
Checkpoint / Orthrus & Val. loss & 40 & -0.022 [-0.036, -0.009] & -0.013 [-0.023, -0.004] \\
Checkpoint / Velox & Final & 40 & -0.008 [-0.017, 0.004] & -0.000 [-0.007, 0.004] \\
Checkpoint / Velox & Val. loss & 40 & -0.011 [-0.025, 0.003] & -0.003 [-0.010, 0.002] \\
Reindexer / Hashed path & Final & 10 & 0.001 [-0.028, 0.035] & -0.011 [-0.024, 0.002] \\
Reindexer / Hashed path & Val. loss & 10 & 0.001 [-0.029, 0.035] & -0.011 [-0.025, 0.003] \\
Reindexer / Type only & Final & 10 & -0.067 [-0.070, -0.061] & -0.077 [-0.080, -0.071] \\
Reindexer / Type only & Val. loss & 10 & -0.014 [-0.035, 0.000] & -0.016 [-0.040, 0.000] \\
Reindexer / word2vec & Final & 40 & -0.059 [-0.082, -0.034] & -0.064 [-0.074, -0.053] \\
Reindexer / word2vec & Val. loss & 40 & -0.051 [-0.073, -0.027] & -0.058 [-0.069, -0.048] \\
\addlinespace[2pt]\multicolumn{5}{l}{\textbf{THEIA E3}} \\
Checkpoint / Orthrus & Final & 40 & -0.012 [-0.019, -0.005] & -0.013 [-0.030, -0.002] \\
Checkpoint / Orthrus & Val. loss & 40 & -0.012 [-0.018, -0.006] & -0.034 [-0.078, -0.007] \\
Checkpoint / Velox & Final & 40 & -0.091 [-0.140, -0.045] & -0.008 [-0.012, -0.004] \\
Checkpoint / Velox & Val. loss & 40 & -0.071 [-0.109, -0.036] & -0.003 [-0.006, -0.001] \\
Reindexer / Hashed path & Final & 10 & 0.058 [-0.016, 0.117] & -0.098 [-0.296, 0.045] \\
Reindexer / Hashed path & Val. loss & 10 & 0.038 [-0.007, 0.085] & -0.177 [-0.481, 0.039] \\
Reindexer / Type only & Final & 10 & -0.092 [-0.098, -0.086] & -0.044 [-0.047, -0.041] \\
Reindexer / Type only & Val. loss & 10 & -0.084 [-0.095, -0.068] & -0.042 [-0.050, -0.033] \\
Reindexer / word2vec & Final & 40 & -0.040 [-0.083, 0.007] & -0.016 [-0.021, -0.011] \\
Reindexer / word2vec & Val. loss & 40 & -0.034 [-0.091, 0.025] & -0.020 [-0.026, -0.014] \\
\bottomrule
\end{tabular}
\caption{Fixed-output target sensitivity: all twenty declared contrasts at the main table's P@100 budget. Checkpoint rows subtract the original Orthrus-test-ADP-selected output from the stated alternative, while reindexer rows subtract released from corrected. Brackets give 95 \% paired-member bootstrap intervals under node-id tie-breaking, not label-interaction tests or tie-order bounds. All scores, alarms and populations are fixed. Only 40 of 12,858 native ThreaTrace CADETS targets and 25,342 of 25,358 THEIA targets occur in these scored populations. This is not evaluation on the complete native population. No alternative-target campaign ADP is defined. All four metrics, silent-pair counts and independent-order tie envelopes remain in the artifact.}\label{tab:target-sensitivity}
\end{table*}

\FloatBarrier

\section{Checkpoint selection and implementation checks}
\textbf{Membership and randomization.} Overlapping expansions retain the larger configuration as a whole, so 74 nonshared members of excluded configurations appear in no table. Principal word2vec comparisons vary featurization seeds at a fixed training seed, while deterministic-feature comparisons vary training seeds. A crossed CADETS diagnostic varies both, but one observation per cell cannot separate interaction from execution noise. Intervals describe randomization on fixed hosts, not new-host generalization.

\textbf{Identical outputs and model identity.} Varying only a featurization seed ignored by the features does not produce independent linear models. Such deterministic copies count once. Conversely, equal predictions from distinct randomized training executions do not justify discarding a member. Training records, not outcome equality alone, must establish the sampling unit and the number of independent repetitions.

\textbf{Generation-consistency exclusions.} The check applies to the detectors thresholded by the max-of-validation rule: Velox, Orthrus (GNN), Kairos and R-CAID\cite{goyal2024rcaid}. The current frozen record has no additional generation-consistency exclusions under either checkpoint rule. This does not establish completeness of the underlying candidate records. The check itself uses no test label. The test-snooped Orthrus variant is not checked: it alerts on the upper of two k-means clusters of its highest unlabelled test scores, so the rule does not produce its saved alerts.

Across the frozen record, minimum validation loss selects epoch 0 for 30 of the 845 Velox and Orthrus evaluation records, seven of them in the forty-member CADETS Orthrus configuration and six in the ten-member type-only CADETS Velox configuration. We report every recoverable, generation-consistent member of each retained configuration (\S2), with its actual sample count, and we treat oracle best-member results as unattainable bounds.

\textbf{Implementation-specific selection rules.} NodLink's released threshold is the 80th percentile of its benign training scores, although its five hyperparameters were chosen on DARPA results. PIDSMaker's NodLink configuration instead uses a validation 90th percentile.

\textbf{Additional common implementations.} Two further families enter our tables only as PIDSMaker's reimplementations, not as audited artifacts. OCR-APT\cite{aly2025ocrapt} performs one-class subgraph anomaly detection. PIDSMaker sets each type's threshold from validation losses, but selects its contamination fraction using the fraction of validation rows whose node identities occur in the configured attack-label set, clipped to the configured lower and upper bounds. This is label contact, although the effect on the saved thresholds has not been established. Its enabled early-stopping helper also uses configured attack identities to evaluate per-type validation F1, AUC or true-negative rate and can freeze encoder and objective submodules. This dependency can therefore affect training as well as thresholds, although its realized effect on the frozen members remains unverified. The training sequence exposes no seed parameter, so its five members differ only through training nondeterminism. R-CAID\cite{goyal2024rcaid} uses root-cause-augmented node embeddings and, in PIDSMaker, takes the max-of-validation threshold. Both take PIDSMaker's test-label checkpoint under the PIDSMaker checkpoint rule, like every integrated system. The complete source-line audit is in the artifact.

\textbf{Checkpoint tie convention.} Both the experimental upstream revision \texttt{ae1e9fd} and the audited PIDSMaker 2.0.0 revision break equal recorded ADP by discrimination. Our reconstruction retains the first maximum instead. Two of the 160 headline members have tied maxima, and the native rule changes one THEIA Orthrus selection, from epoch 5 to 7. Its forty-member mean tie-safe ADP changes by less than $10^{-5}$. Both candidate ADPs round to 0.504, so every three-decimal paired-effect input, and therefore the reported paired mean and interval, is unchanged. This sensitivity covers the four headline configurations, not all reconstructed selections. Neither alternative changes the final or benign-validation-loss selections.

\textbf{Broader checkpoint sensitivity.} Replay evaluates the same 295 members under the final, minimum-validation-loss and recorded test-ADP selectors. Minimum-validation-loss mean ADP is lower than the test-selected mean in 27 of the 30 configurations. The retained paired effects use 2,000-resample intervals. Neither the counts nor these intervals establish independent replications across configurations.
\begin{table*}[tp]\centering\setlength{\tabcolsep}{2pt}\scriptsize
\begin{tabular}{lllc p{0.45\textwidth}}
\toprule
artifact & upstream repository & content commit & selects with test labels & checkpoint; operating point \\
\midrule
Kairos & kairos & \texttt{0e0b633beb46} & opaque & fixed epoch 50; hard-coded per-day queue cutoffs \\
FLASH & Flash-IDS & \texttt{ccbafee7eb79} & no & no best epoch, all 22 curriculum stages; alarm when every stage misclassifies \\
MAGIC & MAGIC & \texttt{aa0b647eea74} & yes & fixed final epoch; threshold from the labelled test PR curve at a target recall \\
ThreaTrace & threaTrace & \texttt{2229474a8b37} & yes & cascade retrained and pruned until test recall > 0.8 and precision > 0.7; fixed probability ratios \\
NodLink & NODLINK & \texttt{434dbe2d0be8} & no & minimum benign training loss; 80th percentile of training scores \\
CAPTAIN & CAPTAIN & \texttt{992f27ad25e5} & no & epoch 99 as released; per-event thresholds learned on benign data \\
ProvCTDG & ProvCTDG & \texttt{5db3e89ececa} & no & validation-metric maximum; fixed score 0.5 \\
Orthrus & orthrus & \texttt{e7f25dfee1dd} & yes & highest test MCC over saved epochs; k-means over the top 20 test scores \\
Velox & PIDSMaker & \texttt{ef9a9e2568a4} & yes & highest test ADP over saved epochs; maximum validation loss \\
PIDSMaker 2.0.0 & PIDSMaker & \texttt{667bc8f9cc7f} & yes & highest test ADP for every integrated system; system-specific thresholds \\
\bottomrule
\end{tabular}
\caption{Source anchors for the selection findings of Section 3: for each released artifact we read, the upstream repository, the audited content commit, whether its selection touches test attack labels, and how the released code chooses its checkpoint and operating point. The audit names the files and repository owners, and Table~\ref{tab:excerpts} quotes the decisive lines. A row marked no is one the audit clears.}\label{tab:anchors}
\end{table*}

\begin{table*}[tp]\centering\setlength{\tabcolsep}{2pt}\scriptsize
\begin{tabular}{>{\raggedright\arraybackslash}p{0.09\textwidth}>{\raggedright\arraybackslash}p{0.12\textwidth}>{\raggedright\arraybackslash}p{0.06\textwidth}>{\raggedright\arraybackslash}p{0.43\textwidth}>{\raggedright\arraybackslash}p{0.22\textwidth}}
\toprule
artifact & component, commit & lines & code & effect \\
\midrule
MAGIC & evaluator, \texttt{aa0b647eea74} & 197, 208, 211 & \texttt{prec, rec, threshold = precision\_recall\_curve(y\_test, score)} / \texttt{if dataset == \textquotesingle{}cadets\textquotesingle{} and rec[i] < 0.9976:} / \texttt{best\_thres = threshold[best\_idx]} & threshold read off the test precision-recall curve at a hard-coded CADETS test recall \\
ThreaTrace & DARPA trainer, \texttt{2229474a8b37} & 95, 138, 255 & \texttt{data, feature\_num, label\_num, adj, adj2, nodeA, \_nodeA, \_neibor = MyDatasetA(path, 0)} / \texttt{if (\_tp/len(nodeA) > 0.8) and (\_tp/(\_tp+\_fp+eps) > 0.7):} / \texttt{flag = validate()} & validate() loads the test graph with its labelled attack nodes (nodeA); the cascade is retrained until test recall exceeds 0.8 and test precision 0.7 \\
Orthrus & evaluator, \texttt{e7f25dfee1dd} & 20, 46 & \texttt{for model\_epoch\_dir in listdir\_sorted(test\_losses\_dir):} / \texttt{if stats["mcc"] > best\_mcc:} & every saved epoch is scored on the test set and the highest test MCC is reported \\
Velox & evaluator, \texttt{ef9a9e2568a4} & 31, 69 & \texttt{for model\_epoch\_dir in sorted\_files:} / \texttt{if stats["adp\_score"] > best\_adp:} & every saved epoch is scored on the test set and the highest test ADP is kept \\
PIDSMaker 2.0.0 & evaluation task, \texttt{667bc8f9cc7f} & 38, 95, 96, 97, 98, 99 & \texttt{for model\_epoch\_dir in sorted\_files:} / \texttt{if best\_model\_selection == "best\_adp":} / \texttt{condition = (stats["adp\_score"] > best\_metrics["adp\_score"]) or (} / \texttt{stats["adp\_score"] == best\_metrics["adp\_score"]} / \texttt{and stats["discrimination"] > best\_metrics["discrimination"]} / \texttt{)} & the same test-ADP epoch loop, ties broken by discrimination, applied to every integrated system \\
PIDSMaker 2.0.0 & Orthrus configuration, \texttt{667bc8f9cc7f} & 97 & \texttt{best\_model\_selection: best\_adp} & the configured rule, set in every integrated system's configuration (Velox inherits it from Orthrus) \\
\bottomrule
\end{tabular}
\caption{Decisive source lines of the test-label selections in Table~\ref{tab:anchors}, verbatim at the audited commits with leading whitespace removed. A spaced slash ( / ) separates quoted lines, and line numbers refer to the named component.}\label{tab:excerpts}
\end{table*}

\begin{table*}[tp]\centering\setlength{\tabcolsep}{2pt}\small
\begin{tabular}{lllll}
\toprule
detector & dataset & n (t/f) & Test-selected ADP [CI] & final ADP [CI] \\
\midrule
Velox (corrected) & CADETS\_E3 & 40/40 & 0.33 [0.26, 0.41] & 0.18 [0.10, 0.27] \\
Velox (default) & CADETS\_E3 & 40/40 & 0.66 [0.61, 0.71] & 0.55 [0.50, 0.61] \\
Orthrus (GNN) & CADETS\_E3 & 40/40 & 0.53 [0.46, 0.60] & 0.28 [0.22, 0.35] \\
MAGIC & CADETS\_E3 & 5/5 & 0.01 0.01--0.02 & 0.01 0.00--0.02 \\
ThreaTrace & CADETS\_E3 & 5/5 & 0.00 0.00--0.00 & 0.00 0.00--0.00 \\
Kairos & CADETS\_E3 & 5/5 & 0.09 0.00--0.33 & 0.01 0.00--0.02 \\
Flash & CADETS\_E3 & 5/5 & 0.34 0.33--0.35 & 0.20 0.00--0.34 \\
NodLink & CADETS\_E3 & 5/5 & 0.59 0.22--0.73 & 0.32 0.00--0.67 \\
OCR-APT & CADETS\_E3 & 5/5 & 0.69 0.68--0.69 & 0.69 0.68--0.69 \\
R-CAID & CADETS\_E3 & 5/5 & 0.19 0.02--0.36 & 0.10 0.01--0.34 \\
Orthrus (test-snooped) & CADETS\_E3 & 5/5 & 0.54 0.39--0.83 & 0.33 0.02--0.63 \\
Velox (corrected) & CADETS\_E3\_CHRONO & 10/10 & 0.67 [0.54, 0.79] & 0.37 [0.19, 0.60] \\
Velox (default) & CADETS\_E3\_CHRONO & 10/10 & 0.88 [0.84, 0.92] & 0.72 [0.57, 0.85] \\
Orthrus (GNN) & CADETS\_E3\_CHRONO & 10/10 & 0.72 [0.59, 0.85] & 0.32 [0.23, 0.40] \\
MAGIC & CADETS\_E3\_CHRONO & 5/5 & 0.04 0.02--0.06 & 0.04 0.02--0.06 \\
ThreaTrace & CADETS\_E3\_CHRONO & 5/5 & 0.01 0.01--0.01 & 0.01 0.01--0.01 \\
Kairos & CADETS\_E3\_CHRONO & 5/5 & 0.13 0.01--0.25 & 0.00 0.00--0.00 \\
Flash & CADETS\_E3\_CHRONO & 5/5 & 0.52 0.50--0.57 & 0.18 0.01--0.50 \\
NodLink & CADETS\_E3\_CHRONO & 5/5 & 0.65 0.02--1.00 & 0.48 0.00--1.00 \\
OCR-APT & CADETS\_E3\_CHRONO & 5/5 & 0.71 0.70--0.72 & 0.71 0.70--0.72 \\
R-CAID & CADETS\_E3\_CHRONO & 5/5 & 0.06 0.02--0.17 & 0.03 0.02--0.04 \\
Orthrus (test-snooped) & CADETS\_E3\_CHRONO & 5/5 & 0.80 0.58--1.00 & 0.24 0.01--0.35 \\
Velox (corrected) & THEIA\_E3 & 40/40 & 0.75 [0.69, 0.82] & 0.42 [0.32, 0.53] \\
Velox (default) & THEIA\_E3 & 40/40 & 0.71 [0.66, 0.76] & 0.42 [0.35, 0.49] \\
Orthrus (GNN) & THEIA\_E3 & 40/40 & 0.57 [0.52, 0.61] & 0.27 [0.21, 0.34] \\
MAGIC & THEIA\_E3 & 5/5 & 0.00 0.00--0.00 & 0.00 0.00--0.00 \\
ThreaTrace & THEIA\_E3 & 5/5 & 0.00 0.00--0.02 & 0.00 0.00--0.00 \\
Kairos & THEIA\_E3 & 5/5 & 0.26 0.25--0.26 & 0.23 0.13--0.26 \\
Flash & THEIA\_E3 & 5/5 & 0.02 0.00--0.07 & 0.00 0.00--0.01 \\
NodLink & THEIA\_E3 & 5/5 & 0.00 0.00--0.00 & 0.00 0.00--0.00 \\
OCR-APT & THEIA\_E3 & 5/5 & 0.12 0.12--0.12 & 0.12 0.12--0.12 \\
R-CAID & THEIA\_E3 & 5/5 & 0.00 0.00--0.01 & 0.00 0.00--0.01 \\
Orthrus (test-snooped) & THEIA\_E3 & 5/5 & 0.37 0.14--0.55 & 0.10 0.02--0.22 \\
Velox (corrected) & CLEARSCOPE\_E3 & 5/5 & 0.22 0.01--0.50 & 0.05 0.01--0.12 \\
Velox (default) & CLEARSCOPE\_E3 & 10/10 & 0.02 [0.02, 0.03] & 0.01 [0.00, 0.01] \\
Orthrus (GNN) & CLEARSCOPE\_E3 & 10/10 & 0.46 [0.39, 0.50] & 0.12 [0.05, 0.21] \\
MAGIC & CLEARSCOPE\_E3 & 5/5 & 0.06 0.04--0.09 & 0.02 0.01--0.04 \\
ThreaTrace & CLEARSCOPE\_E3 & 5/5 & 0.00 0.00--0.01 & 0.00 0.00--0.01 \\
Kairos & CLEARSCOPE\_E3 & 5/5 & 0.50 0.50--0.50 & 0.15 0.01--0.25 \\
Flash & CLEARSCOPE\_E3 & 5/5 & 0.02 0.01--0.03 & 0.00 0.00--0.00 \\
NodLink & CLEARSCOPE\_E3 & 5/5 & 0.02 0.01--0.02 & 0.01 0.01--0.02 \\
OCR-APT & CLEARSCOPE\_E3 & 5/5 & 0.00 0.00--0.00 & 0.00 0.00--0.00 \\
R-CAID & CLEARSCOPE\_E3 & 5/5 & 0.00 0.00--0.00 & 0.00 0.00--0.00 \\
Orthrus (test-snooped) & CLEARSCOPE\_E3 & 5/5 & 0.05 0.02--0.08 & 0.00 0.00--0.01 \\
\bottomrule
\end{tabular}
\caption{Test-selected versus final checkpoints, evaluated by tie-safe ADP. Cells give means [95 \% member-bootstrap intervals] for $n\geq10$ and member ranges otherwise. Counts are executions with saved scores, not independent outcomes. OCR-APT is label-dependent. R-CAID scores different populations on all three E3 hosts and is not rank-comparable. Generation-consistency checks apply to maximum-validation thresholds (Velox, Orthrus GNN, Kairos and R-CAID), not other detectors' native rules. CADETS E3 CHRONO is the chronological variant. --- unavailable.}\label{tab:leaderboard}
\end{table*}

\begin{table*}[tp]\centering\setlength{\tabcolsep}{2pt}\small
\begin{tabular}{lllllll}
\toprule
artifact & 1 Strict & 2 Top-B & 3 Replicated & 4 Independent & 5 Chronological & 6 Transparent \\
\midrule
Kairos & yes$^{a}$ & no$^{b}$ & no$^{c}$ & not established$^{d}$ & yes & no \\
FLASH & no$^{e}$ & no & no$^{f}$ & yes$^{g}$ & yes & no \\
MAGIC & yes$^{h}$ & no & no$^{i}$ & no$^{j}$ & no$^{aa}$ & no \\
ThreaTrace & no$^{k}$ & no & no$^{l}$ & no$^{m}$ & yes & no \\
NodLink & partial$^{n}$ & no$^{o}$ & no$^{p}$ & partial$^{q}$ & yes & no \\
CAPTAIN & yes$^{r}$ & no & no$^{s}$ & yes$^{t}$ & yes & no \\
Reha et al. (ProvCTDG) & yes$^{u}$ & no & partial$^{v}$ & yes$^{w}$ & yes & no \\
Orthrus & yes & no$^{x}$ & partial$^{y}$ & no$^{z}$ & no$^{A}$ & no \\
Velox & yes & no$^{B}$ & partial$^{C}$ & no$^{D}$ & no$^{E}$ & no \\
PIDSMaker 2.0.0 (all systems) & yes & no & partial$^{F}$ & no$^{G}$ & no$^{H}$ & partial$^{I}$ \\
STRICT evaluation (ours) & yes & yes$^{J}$ & yes$^{K}$ & partial$^{L}$ & partial$^{M}$ & partial$^{N}$ \\
\bottomrule
\end{tabular}
\caption{STRICT checklist. Which of the six rules each released artifact's evaluation path satisfies (from the source audit of \S3). The last row is our STRICT evaluation, assessed by the same rules, with inherited hyperparameter provenance unverified (Rule 4 partial), chronology partial and public selection evidence incomplete (Rule 6 partial). The rules are not points in a score. Rule 1 permits exact native targets: nodes (MAGIC), windows (Kairos), type/name entities (CAPTAIN) and edges (ProvCTDG). NodLink is partial because process exports and graph-recall credit differ. Selection is assessed separately under Rule 4. Lettered evidence notes appear directly below.}\label{tab:checklist}

\par\medskip\begin{minipage}{\textwidth}\footnotesize a: Kairos uses exact matching at native time-window filenames, and an alerted queue marks its member windows. b: Kairos uses hard-coded per-day queue cutoffs. c: Kairos has one model. d: Kairos uses hard-coded constants of undocumented provenance, so independence is not established rather than violated. e: FLASH uses 2-hop relaxation in the released notebooks. f: FLASH has one released model per dataset. g: FLASH uses benign validation. h: MAGIC uses exact native node labels, with test-label threshold selection assessed under Rule 4. i: MAGIC has one checkpoint. j: MAGIC chooses its threshold from test labels. k: ThreaTrace uses 2-hop relaxation and a 12,858-node label set, of which 12,852 resolve in the test shard. l: ThreaTrace reports one cascade. m: ThreaTrace accepts its cascade on test recall and precision.

n: NodLink exports exact filtered-process membership, but cached-graph recall also credits processes through the tailoring map. o: NodLink uses the 80th percentile of benign training scores. p: NodLink has one checkpoint. q: NodLink's threshold rule is clean, although its hyperparameters were chosen on DARPA results. r: CAPTAIN uses exact native type/name entity matching and reports zero-hop and one-hop targets separately. s: CAPTAIN uses epoch 99 of one model. t: CAPTAIN uses benign-trained tau values. u: Reha et al. (ProvCTDG) use exact native edge-hash matching. v: Reha et al. (ProvCTDG) report the best validation result over several executions. w: Reha et al. (ProvCTDG) use a validation metric.

x: Orthrus replaces max-validation with top-20 test-score clustering. y: Orthrus reports the best of 5 seeds. z: Orthrus selects the epoch by test MCC. A: Orthrus's CADETS attack day precedes training. B: Velox uses the maximum of one validation day. C: Velox reports the best of 5 seeds. D: Velox selects the epoch by test ADP. E: Velox, CADETS. F: PIDSMaker 2.0.0 reports the best of 5 seeds for all systems. G: PIDSMaker 2.0.0 selects the epoch by test ADP for every system. H: PIDSMaker 2.0.0, CADETS for all systems. I: PIDSMaker 2.0.0 releases the framework without score files for any system.

J: declared ranking budgets and exact within-tie coverage bounds. K: declared randomization and member counts, with member-wise scores retained. The printed ADP summary uses bootstrap intervals for at least ten members and observed ranges otherwise. Native single-checkpoint reproductions do not establish training variability. L: checkpoint selection is test-independent, but inherited hyperparameter-selection provenance remains unverified. M: configured CADETS is retrospective, with chronological comparisons identified separately. N: scores and code are public, but threshold-reconstruction inputs cover only the headline configurations. aa: MAGIC's training data include a later shard and omit edges incident to labelled nodes (\S3.1).

Release note for N: the score record and numerical validation arrays are public on \href{https://huggingface.co/datasets/STRICT-APT/strict-score-record}{Hugging Face}, with anonymous payload retrieval matching the recorded hashes. The \href{https://zenodo.org/records/22990291}{Zenodo metadata record (DOI 10.5281/zenodo.22990291)} archives hashes, member records and validation arrays, not the score files themselves. The \href{https://zenodo.org/records/22991158}{Zenodo code record (DOI 10.5281/zenodo.22991158)} supplies the evaluation implementation and analysis code, and its anonymous download matches the recorded hash. N remains partial because independent threshold reconstruction outside the four headline configurations still needs unreleased validation losses.\end{minipage}
\end{table*}

\FloatBarrier

\section{Reindexer controls}
\textbf{Degree-dependent feature scaling.} The main-text derivation assumes copies of the same static node vector in both roles. Nodes present in only one role retain that vector. Type-only Velox consequently produces 2,151--2,330 distinct CADETS node scores per member over its ten members, which is 0.76--0.83 \% of 281,585 test nodes, although its three node types and ten edge types allow at most 90 distinct (source type, destination type, edge type) losses. Orthrus is not affected, because its released encoder keeps distinct source and destination representations, following TGN's separate source and destination message functions\cite{rossi2020tgn}. Default-reindexer Velox outcomes therefore remain evidence about the released PIDSMaker setting rather than a clean architecture effect.

\textbf{Identical-input mechanism check.} A separate execution feeds cloned tensors from the lexicographically first retained CADETS E3 test window through both reindexer branches, selected before comparing their outputs. All 4,407 edges across five configured batches are included. Both branches match independently accumulated embedding and node-type formulas, with maximum absolute error below $8.1\times10^{-7}$. Outputs differ at 132 of 1,056 node--batch occurrences. This verifies the transformation of identical current inputs, not historical feature identity or the retraining performance differences.

\textbf{Historical paired sensitivity.} Table~\ref{tab:pairs} reports both checkpoint rules. Paired ADP values correlate weakly (Pearson $|r|\le 0.15$ on the four headline word2vec host/checkpoint contrasts), and paired and unpaired intervals have similar widths. Matching seeds adds little precision to these estimates.

\textbf{Earlier identical-input control.} Three CADETS type-only initialization pairs fix realized inputs and initial states. Corrected-minus-released ADP is $-0.046$, $-0.056$ and $-0.522$ (mean $-0.208$), with range $[-0.522,-0.046]$. This supports a mechanism effect on retained inputs, not recovery of historical feature realizations. All six outcomes remain in the experimental record.

The native factory executes dropout 0.0 despite configured dropout 0.3 in this earlier experiment. The discrepancy was identified after the first pair's outcomes were known. The original design and all six outcomes are retained without replacement. Correction raises native-threshold false positives from 78 to 1,079 in every pair. This deviation is separate from the verified-dropout controls below.

\textbf{Verified-dropout replication.} A subsequent design fixed five initialization pairs on each of CADETS E3 and THEIA E3 before preparing or training these members, but after the earlier CADETS outcomes and historical host comparisons were known. It is not an outcome-blind test on unseen benchmarks. Both conditions use actual and configured dropout 0.3, set by the same scoped adapter to the existing dropout module. The declared initializations are 0 through 4, released then corrected within each pair, with twelve epochs, width 256, learning rate $10^{-4}$ and fixed selection of epoch 11. Both conditions consume identical complete pre-reindexer batches and begin with matched parameters, optimizer state and random states. Their realized architectures are checked separately.

Both hosts use type-only inputs and their configured splits. CADETS projects endpoint and event types from retained messages, while THEIA regenerates its batches from retained graphs using the native type-only producer. Inputs and target mappings are bound before training. Table~\ref{tab:matched-replication} gives every paired outcome, without pooling hosts or the earlier dropout-0.0 experiment. Numerical verification reconstructs all twenty members' scores, validation thresholds, alerts and metrics from retained losses.

THEIA's released-condition ADP repeats for initializations 1 and 2, and for 3 and 4 to numerical precision. Their validation thresholds differ, and corrected ADPs also differ slightly. Thus five trainings give three distinct released ADP outcomes, not five distinct ranking summaries. All pairs remain in the record and the reported mean.

Ranking and threshold effects differ. Every corrected CADETS member raises 2 true and 1,079 false positives, whereas four corrected THEIA members raise no alert and the fifth raises 21 false positives without a true positive. Precision for a silent member is undefined, not zero (Table~\ref{tab:matched-replication}). Removing degree-dependent scaling therefore neither uniformly increases false positives nor establishes improved detection effectiveness. These are conditional mechanism tests, not semantic label validation, raw-log reproduction or population-level generalization.

\textbf{Historical corrected controls.} Corrected CADETS word2vec has a heavy false-alert tail: four of forty members raise more than a thousand alerts each, while the remainder average fewer than four. Pooled and mean-member precision therefore differ. Strong hashed-path ranking does not separate general attack structure from benchmark-specific names.
\begin{table*}[tp]\centering\setlength{\tabcolsep}{2pt}\small
\begin{tabular}{lllrr}
\toprule
Features & Checkpoint & Pairs & ADP $d$ paired & P@100 $d$ paired \\
\midrule
\addlinespace[2pt]\multicolumn{5}{l}{\textbf{CADETS E3}} \\
Hashed path & final & 10 & +0.015 [+0.007, +0.024] & +0.001 [-0.029, +0.035] \\
Hashed path & val. loss & 10 & +0.015 [+0.007, +0.024] & +0.001 [-0.029, +0.034] \\
Type only & final & 10 & -0.385 [-0.532, -0.235] & -0.067 [-0.070, -0.061] \\
Type only & val. loss & 10 & -0.017 [-0.031, -0.004] & -0.014 [-0.035, +0.000] \\
word2vec & final & 40 & -0.376 [-0.472, -0.269] & -0.059 [-0.081, -0.035] \\
word2vec & val. loss & 40 & -0.294 [-0.391, -0.197] & -0.051 [-0.073, -0.027] \\
\addlinespace[2pt]\multicolumn{5}{l}{\textbf{CLEARSCOPE E3}} \\
word2vec & final & 5 & +0.049 [+0.006, +0.094] & +0.008 [+0.004, +0.010] \\
word2vec & val. loss & 5 & +0.015 [+0.007, +0.024] & +0.008 [+0.004, +0.010] \\
\addlinespace[2pt]\multicolumn{5}{l}{\textbf{CLEARSCOPE E5}} \\
word2vec & final & 5 & +0.082 [-0.006, +0.189] & +0.020 [-0.008, +0.050] \\
word2vec & val. loss & 5 & +0.123 [-0.006, +0.246] & +0.046 [+0.008, +0.078] \\
\addlinespace[2pt]\multicolumn{5}{l}{\textbf{OpTC-051}} \\
word2vec & final & 5 & -0.003 [-0.016, +0.010] & -0.004 [-0.010, +0.004] \\
word2vec & val. loss & 5 & -0.003 [-0.020, +0.015] & -0.000 [-0.022, +0.018] \\
\addlinespace[2pt]\multicolumn{5}{l}{\textbf{OpTC-201}} \\
word2vec & final & 5 & +0.189 [+0.017, +0.416] & +0.052 [+0.002, +0.130] \\
word2vec & val. loss & 5 & -0.088 [-0.239, +0.024] & +0.004 [-0.022, +0.032] \\
\addlinespace[2pt]\multicolumn{5}{l}{\textbf{OpTC-501}} \\
word2vec & final & 5 & -0.037 [-0.087, +0.007] & -0.004 [-0.010, +0.004] \\
word2vec & val. loss & 5 & -0.007 [-0.138, +0.132] & +0.008 [-0.004, +0.020] \\
\addlinespace[2pt]\multicolumn{5}{l}{\textbf{THEIA E3}} \\
Hashed path & final & 10 & +0.065 [-0.142, +0.264] & +0.058 [-0.017, +0.121] \\
Hashed path & val. loss & 10 & +0.084 [-0.160, +0.284] & +0.038 [-0.006, +0.087] \\
Type only & final & 10 & -0.453 [-0.526, -0.385] & -0.092 [-0.097, -0.085] \\
Type only & val. loss & 10 & -0.477 [-0.559, -0.404] & -0.084 [-0.095, -0.068] \\
word2vec & final & 40 & +0.006 [-0.112, +0.115] & -0.040 [-0.087, +0.005] \\
word2vec & val. loss & 40 & -0.007 [-0.128, +0.121] & -0.034 [-0.091, +0.020] \\
\addlinespace[2pt]\multicolumn{5}{l}{\textbf{THEIA E5}} \\
word2vec & final & 5 & +0.145 [-0.087, +0.429] & +0.004 [-0.008, +0.016] \\
word2vec & val. loss & 5 & +0.009 [-0.103, +0.102] & +0.008 [-0.008, +0.020] \\
\bottomrule
\end{tabular}
\caption{Historical reindexer contrasts: corrected minus default, paired by recorded featurization and training seeds. Means [95 \% bootstrap intervals over pairs], conditional on the recorded settings. Seed matching does not establish identical realized inputs. Five-pair intervals are exploratory. Native-threshold, unpaired and correlation diagnostics remain in the retained record.}\label{tab:pairs}
\end{table*}

\begin{table*}[tp]\centering\setlength{\tabcolsep}{6pt}\small
\begin{tabular}{lrrrrr}
\toprule
Initialization & Released ADP & Corrected ADP & $\Delta$ADP & Released TP / FP & Corrected TP / FP \\
\midrule
\addlinespace[2pt]\multicolumn{6}{l}{\textbf{CADETS E3}} \\
0 & 0.459658 & 0.001511 & -0.458147 & 7 / 78 & 2 / 1,079 \\
1 & 0.057443 & 0.001474 & -0.055968 & 7 / 78 & 2 / 1,079 \\
2 & 0.043546 & 0.001269 & -0.042277 & 0 / 77 & 2 / 1,079 \\
3 & 0.063208 & 0.001201 & -0.062007 & 7 / 78 & 2 / 1,079 \\
4 & 0.475933 & 0.001250 & -0.474684 & 7 / 78 & 2 / 1,079 \\
\addlinespace[2pt]\multicolumn{6}{l}{\textbf{THEIA E3}} \\
0 & 0.417998 & 0.000531 & -0.417467 & 0 / 2 & 0 / 21 \\
1 & 0.749916 & 0.000533 & -0.749383 & 2 / 2 & 0 / 0 \\
2 & 0.749916 & 0.000531 & -0.749385 & 2 / 2 & 0 / 0 \\
3 & 0.659782 & 0.000531 & -0.659251 & 0 / 2 & 0 / 0 \\
4 & 0.659782 & 0.000532 & -0.659250 & 0 / 2 & 0 / 0 \\
\bottomrule
\end{tabular}
\caption{Identical-input type-only controls: all five initialization pairs per host, fixed epoch 11, actual and configured dropout 0.3. THEIA has five trainings but three distinct released ADP outcomes. $\Delta$ADP is corrected minus released. TP / FP uses each condition's maximum-validation threshold. 0 / 0 denotes silence, with undefined precision. These individual outcomes are not confidence intervals and remain separate from the earlier dropout-0.0 experiment.}\label{tab:matched-replication}
\end{table*}

\FloatBarrier

\section{Ranking and calibration}
\textbf{Native-window threshold sensitivity.} A separate Kairos case preserves its native window unit. With one released checkpoint and identical queue scores, a fixed reference cutoff of 100 yields 4 TP and 1 FP (precision 0.800), whereas the benign-validation maximum of 5.037 yields 4 TP and 38 FP (precision 0.0952), both with recall 1 over 179 test windows. The reference cutoff matches PIDSMaker's constant, not a reconstruction of the native artifact's undocumented day-specific cutoffs. Execution uses compatibility repairs around the released weights. This is a threshold-sensitivity case, not a repeated-training estimate, a native-paper-row reproduction, or a node-level leaderboard comparison.

\textbf{Validation-size diagnostic.} Figure~\ref{fig:valsize} fixes one final-checkpoint CADETS member and subsamples its 111,331-node validation day to 500 nodes over 100 draws. Median false positives rise from 2 to 464, with interquartile range 203--869. The CLEARSCOPE diagnostic fixes a different member: validation contains 70 netflows, 61 unseen during training, and 8,496 of its 10,340 false alerts are unseen netflows. These are single-member mechanism illustrations.

\textbf{Checkpoint comparisons.} Table~\ref{tab:leaderboard} compares test-selected and final-checkpoint ADP. Its intervals describe member variation, not a common ranking across unequal populations. Threshold-dependent TP, FP and pooled precision remain in the experimental record.

\textbf{Coverage contrasts with the supervision diagnostic.} On THEIA, every recorded Kairos member has higher ADP but a larger full-coverage cost $B^{*}$ than every OCR-APT member under both checkpoint rules and every within-score tie order. This counterexample does not depend on averaging members or on choosing a favourable tie-break. It concerns the recorded score orderings, not a fair comparison of unsupervised training procedures. On THEIA, corrected Velox's mean ADP is 0.423, against OCR-APT's 0.123. Yet Velox reaches both campaigns within 100 ranked nodes in 15 of 40 members, versus 5 of 5 for OCR-APT. Orthrus reaches both in 3 of 40 members and Kairos in 0 of 5.

This is not simply a difference in the number of true positives: P@100 is 0.099 for Velox and 0.040 for OCR-APT. Finding more labelled nodes need not find more distinct campaigns. OCR-APT's five final members have identical reported outcomes here, so they constitute one observed outcome, not five independent confirmations. The small member sets and fixed hosts limit the inference to these comparisons, so the curves do not estimate deployment reliability.

CADETS retains the non-reversal: OCR-APT leads these configurations on ADP and covers all campaigns within 13 ranked nodes in every recorded member.

\textbf{Coverage-comparison sensitivity.} Under minimum-validation selection, Velox reaches both campaigns within 100 nodes in 13 of 40 members, against 5 of 5 for OCR-APT. The comparison holds the node population, labels and campaign mapping fixed across the four configurations. The figure's bands are exact attainable tie-order bounds, not confidence intervals over independent trials. All final-checkpoint THEIA full-coverage decisions at $B=100$ are invariant to tie order. The complete per-member scores and both checkpoint rules are retained for recomputation. These are offline rankings over the complete test period, not a causal policy accepting 100 alerts per day.

\textbf{Extended comparisons.} Table~\ref{tab:budget} adds configurations absent from the main table, while Table~\ref{tab:edr} covers ATLASv2 and Carbanak. Ranking and thresholded alerts measure different quantities: large false-positive counts can coexist with useful ordering, and rounded-zero precision need not mean zero detections. Native results remain separate because preprocessing, populations, splits and operating points differ. CAPTAIN's binary alarms do not define a within-alert ranking.

\textbf{Exact within-tie coverage bounds.} For each campaign, find its highest-scored labelled node. The lowest of these scores defines the last tie block needed for full coverage. Let $H$ nodes score above it, and let the block contain $M$ nodes. Among campaigns not covered above it, let $c_j$ count each campaign's labelled nodes in the block. The worst ordering gives $B^{*}_{\max}=H+M-\min_j c_j+1$. The best gives $B^{*}_{\min}=H+\tau$, where $\tau$ is the smallest number of block nodes covering those campaigns, found by solving set cover exactly because one node can carry labels of several campaigns. Averaging the per-member endpoints gives attainable bounds on the configuration mean. The released member-level bounds permit other summaries.

\textbf{Heavy tails in full-coverage cost.} Each member's $B^{*}$ is the largest of its campaigns' first-detection ranks, so one late campaign can dominate a configuration mean when there are few campaigns. CADETS holds three: one Orthrus CADETS member needs 38,067 alerts, 952 of the configuration's mean of 1,228, while the other 39 average 283. Per-member values should accompany the mean.
\begin{table*}[tp]\centering\setlength{\tabcolsep}{2pt}\small
\begin{tabular}{lrrrr}
\toprule
Configuration & $n$ & ADP & P@100 & $B^{*}$ \\
\midrule
\addlinespace[2pt]\multicolumn{5}{l}{\textbf{CADETS E3}} \\
v-hfh10-cadets & 10 & 0.98 [0.98, 0.99] & 0.26 & 15 \\
v-hfhfix10-cadets & 10 & 1.00 [1.00, 1.00] & 0.26 & 11 \\
v-type10-cadets & 10 & 0.39 [0.23, 0.53] & 0.07 & 1,276 \\
v-typefix10-cadets $^{\dagger}$ & 10 & 0.00 [0.00, 0.00] & 0.00 & 46,997 \\
v-w2v40-cadets $^{\dagger}$ & 40 & 0.55 [0.50, 0.61] & 0.11 & 263 \\
\addlinespace[2pt]\multicolumn{5}{l}{\textbf{CADETS E3 (chronological)}} \\
v-w2vfix-cadetsC $^{\dagger}$ & 10 & 0.37 [0.19, 0.60] & 0.10 & 152 \\
\addlinespace[2pt]\multicolumn{5}{l}{\textbf{CLEARSCOPE E3}} \\
v-w2v10-clearscope $^{\dagger}$ & 10 & 0.01 [0.00, 0.01] & 0.00 & 200 \\
\addlinespace[2pt]\multicolumn{5}{l}{\textbf{CLEARSCOPE E5}} \\
o-w2v-clearscope5 $^{\dagger}$ & 5 & 0.24 [0.17, 0.25] & 0.07 & 77,927 \\
v-w2vfix-clearscope5 $^{\dagger}$ & 5 & 0.29 [0.22, 0.32] & 0.09 & 7,337 \\
\addlinespace[2pt]\multicolumn{5}{l}{\textbf{OpTC-051}} \\
v-w2vfix-optc051 $^{\dagger}$ & 5 & 0.02 [0.01, 0.05] & 0.01 & 95 \\
\addlinespace[2pt]\multicolumn{5}{l}{\textbf{OpTC-201}} \\
v-w2v10-optc201 $^{\dagger}$ & 10 & 0.29 [0.19, 0.39] & 0.03 & 5 \\
v-w2vfix-optc201 $^{\dagger}$ & 5 & 0.49 [0.17, 0.78] & 0.08 & 6 \\
\addlinespace[2pt]\multicolumn{5}{l}{\textbf{OpTC-501}} \\
v-w2vfix-optc501 $^{\dagger}$ & 5 & 0.10 [0.01, 0.25] & 0.01 & 60 \\
\addlinespace[2pt]\multicolumn{5}{l}{\textbf{THEIA E3}} \\
v-hfh10-theia $^{\dagger}$ & 10 & 0.22 [0.04, 0.44] & 0.04 & 515 \\
v-hfhfix10-theia $^{\dagger}$ & 10 & 0.28 [0.14, 0.42] & 0.10 & 354 \\
v-typefix10-theia $^{\dagger}$ & 10 & 0.00 [0.00, 0.00] & 0.00 & 26,891 \\
v-w2v40-theia $^{\dagger}$ & 40 & 0.42 [0.35, 0.49] & 0.14 & 278 \\
\addlinespace[2pt]\multicolumn{5}{l}{\textbf{THEIA E5}} \\
o-w2v-theia5 $^{\dagger}$ & 5 & 0.00 [0.00, 0.00] & 0.00 & 2,124 \\
v-w2vfix-theia5 & 5 & 0.27 [0.01, 1.00] & 0.02 & 65 \\
\bottomrule
\end{tabular}
\caption{Additional final-checkpoint configurations, grouped by host. Rows already in Table~\ref{tab:strict-results} are not repeated. ADP gives a 95 \% member-bootstrap interval for at least ten members and the observed range otherwise. P@100 and full-coverage cost are means under the same strict targets. $^{\dagger}$ marks tie-sensitive coverage cost. Default-reindexer configurations are diagnostics. The chronological CADETS variant is identified separately.}\label{tab:budget}
\end{table*}

\begin{table*}[tp]\centering\setlength{\tabcolsep}{4pt}\small
\begin{tabular}{lrrr}
\toprule
Configuration & Affected / $n$ & Node-id mean $B^{*}$ & Attainable mean $[B^{*}_{\min}, B^{*}_{\max}]$ \\
\midrule
\addlinespace[2pt]\multicolumn{4}{l}{\textbf{CADETS E3}} \\
o-w2v40-cadets & 3 / 40 & 1,227.67 & [1,227.65, 1,227.72] \\
v-typefix10-cadets & 10 / 10 & 46,997.00 & [9,249.00, 64,879.60] \\
v-w2v40-cadets & 3 / 40 & 263.30 & [263.25, 263.32] \\
v-w2vfix-cadets & 10 / 40 & 1,173.60 & [1,167.15, 1,184.17] \\
x-flash-cadets & 4 / 5 & 40,826.00 & [40,820.80, 40,844.00] \\
x-kairos-cadets & 2 / 5 & 8,804.20 & [8,801.20, 8,804.20] \\
\addlinespace[2pt]\multicolumn{4}{l}{\textbf{CADETS E3 (chronological)}} \\
v-w2vfix-cadetsC & 8 / 10 & 152.10 & [117.00, 217.80] \\
\addlinespace[2pt]\multicolumn{4}{l}{\textbf{CLEARSCOPE E3}} \\
o-w2v10-clearscope & 10 / 10 & 15.30 & [14.30, 15.30] \\
v-w2v10-clearscope & 2 / 10 & 199.60 & [199.40, 199.60] \\
v-w2vfix-clearscope & 4 / 5 & 47.40 & [43.80, 56.20] \\
x-flash-clearscope & 3 / 5 & 11,311.20 & [11,300.40, 11,312.00] \\
x-kairos-clearscope & 5 / 5 & 46.00 & [45.00, 46.00] \\
x-magic-clearscope & 4 / 5 & 1,632.20 & [1,122.20, 1,753.60] \\
x-orthrus\_snooped-clearscope & 4 / 5 & 351.60 & [335.60, 351.60] \\
x-rcaid-clearscope & 5 / 5 & 2,328.00 & [761.20, 8,250.60] \\
x-threatrace-clearscope & 5 / 5 & 15,805.60 & [15,794.40, 15,877.80] \\
\addlinespace[2pt]\multicolumn{4}{l}{\textbf{CLEARSCOPE E5}} \\
o-w2v-clearscope5 & 4 / 5 & 77,926.80 & [72,550.80, 133,634.80] \\
v-w2vfix-clearscope5 & 3 / 5 & 7,336.60 & [7,336.60, 7,337.20] \\
\addlinespace[2pt]\multicolumn{4}{l}{\textbf{OpTC-051}} \\
v-w2vfix-optc051 & 1 / 5 & 95.40 & [95.20, 95.40] \\
\addlinespace[2pt]\multicolumn{4}{l}{\textbf{OpTC-201}} \\
v-w2v10-optc201 & 10 / 10 & 4.70 & [4.70, 5.80] \\
v-w2vfix-optc201 & 5 / 5 & 6.40 & [6.20, 15.60] \\
\addlinespace[2pt]\multicolumn{4}{l}{\textbf{OpTC-501}} \\
v-w2vfix-optc501 & 4 / 5 & 59.80 & [59.60, 60.40] \\
\addlinespace[2pt]\multicolumn{4}{l}{\textbf{THEIA E3}} \\
o-w2v40-theia & 15 / 40 & 15,377.10 & [15,376.77, 15,377.35] \\
v-hfh10-theia & 1 / 10 & 515.30 & [515.30, 515.40] \\
v-hfhfix10-theia & 1 / 10 & 354.20 & [353.80, 354.20] \\
v-typefix10-theia & 10 / 10 & 26,891.30 & [23.90, 46,639.30] \\
v-w2v40-theia & 17 / 40 & 277.77 & [277.77, 278.60] \\
v-w2vfix-theia & 28 / 40 & 442.75 & [442.75, 444.48] \\
x-flash-theia & 3 / 5 & 3,002.60 & [2,952.40, 3,149.60] \\
x-kairos-theia & 2 / 5 & 1,599.80 & [1,599.80, 1,600.20] \\
x-orthrus\_snooped-theia & 2 / 5 & 175.40 & [174.60, 175.40] \\
\addlinespace[2pt]\multicolumn{4}{l}{\textbf{THEIA E5}} \\
o-w2v-theia5 & 4 / 5 & 2,123.80 & [789.40, 3,198.40] \\
\bottomrule
\end{tabular}
\caption{Sensitivity of full-campaign-coverage cost to within-score-tie ordering. Every $^{\dagger}$ configuration in Table~\ref{tab:strict-results} and Table~\ref{tab:budget} appears here. Unmarked measured configurations have identical bounds. The attainable endpoints are means of exact per-member minima and maxima, rounded to two decimals, not confidence intervals. The affected count is the number of members whose endpoints differ. The original node-id means and bootstrap intervals are unchanged. Only the order within equal-score blocks varies. The artifact gives all 133 configurations and member-level bounds.}\label{tab:bstar-ties}
\end{table*}

\begin{table*}[tp]\centering\setlength{\tabcolsep}{2pt}\scriptsize
\begin{tabular}{lllll}
\toprule
detector & dataset & n (t/f) & Test-selected TP / FP / $P_{\mathrm{pool}}$ / ADP [CI] & final TP / FP / $P_{\mathrm{pool}}$ / ADP [CI] \\
\midrule
Velox (corrected) & ATLASV2\_EDR & 10/10 & 0.0 / 0.0 / 0.00 / 0.04 [0.03, 0.04] & 0.0 / 1.1 / 0.00 / 0.01 [0.01, 0.01] \\
Velox (default) & ATLASV2\_EDR & 10/10 & 0.0 / 0.0 / 0.00 / 0.03 [0.03, 0.04] & 0.0 / 2.8 / 0.00 / 0.01 [0.01, 0.01] \\
Orthrus (GNN) & ATLASV2\_EDR & 10/10 & 0.3 / 35 / 0.01 / 0.02 [0.02, 0.02] & 0.7 / 30 / 0.02 / 0.02 [0.01, 0.02] \\
MAGIC & ATLASV2\_EDR & 5/5 & 40.0 / 89422 / 0.00 / 0.02 0.01--0.03 & 40.0 / 90586 / 0.00 / 0.01 0.00--0.03 \\
ThreaTrace & ATLASV2\_EDR & 5/5 & 48.0 / 152342 / 0.00 / 0.00 0.00--0.00 & 48.0 / 152599 / 0.00 / 0.00 0.00--0.00 \\
Kairos & ATLASV2\_EDR & 5/5 & 1.0 / 12 / 0.08 / 0.10 0.04--0.15 & 0.0 / 22 / 0.00 / 0.02 0.01--0.04 \\
Flash & ATLASV2\_EDR & 5/5 & 4.2 / 5447 / 0.00 / 0.00 0.00--0.00 & 14.6 / 46874 / 0.00 / 0.00 0.00--0.00 \\
NodLink & ATLASV2\_EDR & 5/5 & 13.6 / 21338 / 0.00 / 0.00 0.00--0.00 & 10.8 / 22227 / 0.00 / 0.00 0.00--0.00 \\
OCR-APT & ATLASV2\_EDR & 5/5 & 4.0 / 1411 / 0.00 / 0.00 0.00--0.00 & 1.0 / 1075 / 0.00 / 0.00 0.00--0.00 \\
R-CAID & ATLASV2\_EDR & 5/5 & 0.0 / 21 / 0.00 / 0.00 0.00--0.00 & 0.0 / 225 / 0.00 / 0.00 0.00--0.00 \\
Orthrus (test-snooped) & ATLASV2\_EDR & 5/5 & 0.6 / 14 / 0.04 / 0.02 0.01--0.04 & 0.0 / 8.2 / 0.00 / 0.01 0.00--0.02 \\
Velox (corrected) & CARBANAKV2\_EDR & 10/10 & 0.0 / 11 / 0.00 / 0.02 [0.00, 0.05] & 0.0 / 1.9 / 0.00 / 0.01 [0.00, 0.01] \\
Velox (default) & CARBANAKV2\_EDR & 10/10 & 0.0 / 58 / 0.00 / 0.03 [0.00, 0.06] & 0.0 / 3.7 / 0.00 / 0.02 [0.00, 0.04] \\
Orthrus (GNN) & CARBANAKV2\_EDR & 10/10 & 0.0 / 4.9 / 0.00 / 0.02 [0.01, 0.03] & 0.0 / 9.0 / 0.00 / 0.00 [0.00, 0.00] \\
MAGIC & CARBANAKV2\_EDR & 5/5 & 92.6 / 903444 / 0.00 / 0.02 0.00--0.05 & 75.4 / 914789 / 0.00 / 0.01 0.00--0.02 \\
ThreaTrace & CARBANAKV2\_EDR & 5/5 & 215.0 / 1467961 / 0.00 / 0.00 0.00--0.00 & 215.0 / 1467961 / 0.00 / 0.00 0.00--0.00 \\
Kairos & CARBANAKV2\_EDR & 5/5 & 0.0 / 4.2 / 0.00 / 0.00 0.00--0.01 & 0.0 / 2.8 / 0.00 / 0.00 0.00--0.01 \\
Flash & CARBANAKV2\_EDR & 5/5 & 31.6 / 148107 / 0.00 / 0.00 0.00--0.00 & 82.6 / 585041 / 0.00 / 0.00 0.00--0.00 \\
NodLink & CARBANAKV2\_EDR & 5/5 & 112.0 / 186785 / 0.00 / 0.00 0.00--0.00 & 97.6 / 183545 / 0.00 / 0.00 0.00--0.00 \\
OCR-APT & CARBANAKV2\_EDR & 5/5 & 12.0 / 927 / 0.01 / 0.00 0.00--0.00 & 0.0 / 4853 / 0.00 / 0.00 0.00--0.00 \\
R-CAID & CARBANAKV2\_EDR & 5/5 & 0.0 / 4.8 / 0.00 / 0.02 0.01--0.02 & 0.0 / 3.2 / 0.00 / 0.00 0.00--0.00 \\
Orthrus (test-snooped) & CARBANAKV2\_EDR & 5/5 & 0.0 / 17 / 0.00 / 0.01 0.00--0.01 & 0.0 / 16 / 0.00 / 0.00 0.00--0.00 \\
\bottomrule
\end{tabular}
\caption{Endpoint detection and response datasets beyond E3: ATLASv2 and Carbanak v2. Cells give mean TP / mean FP / pooled precision / tie-safe ADP. Uncertainty follows Table~\ref{tab:leaderboard}. Carbanak R-CAID omits 18 of the other systems' 215 labelled targets and scores a different node population, so it is not rank-comparable. --- unavailable.}\label{tab:edr}
\end{table*}

\FloatBarrier

\section{Chronological capacity and padding}
\textbf{Temporal labels and their validation.} For temporal labels, we take each statically labelled node's first incident raw-database event inside its configured half-open attack interval. This reference is detector-independent, but reuses Orthrus node identities and attack windows. Agreement therefore checks timestamp construction rather than malicious-event semantics. ThreaTrace's released label lists\cite{wang2022threatrace} lack timestamps and campaign IDs, and Reha et al.'s released implementation (ProvCTDG)\cite{reha2023provctdg} supplies a partial THEIA\_E3 source-edge cross-check. These checks establish source consistency, not complete semantic validity, particularly for THEIA\_E5.

\textbf{Chronological split sensitivity.} Configured CADETS E3 tests an attack day that precedes some training days, so it supports retrospective ranking only. Removing that test day yields a chronological variant but also removes a campaign. Its Velox/Orthrus mean ordering depends on checkpoint selection and its intervals overlap throughout (Table~\ref{tab:leaderboard}). This does not isolate a chronology effect. THEIA E3/E5 and CADETS E5 support the chronological policy replays.

\textbf{Capacity-policy specification.} Each node's score is its maximum incident-edge loss in a fixed two-hour UTC block. The corrected THEIA comparisons use the first five declared members and a two-of-five vote, not outcome-selected members. For validation exposure $D_v$ host-days and $N_i$ node-block pairs, member $i$ takes the empirical upper quantile at $p_i=\min(bD_v/N_i,1)$. For ascending scores indexed from zero, $Q^{\mathrm{higher}}_{1-p_i}=s_{\lceil(N_i-1)(1-p_i)\rceil}$. Forty bisection steps fit the marginal budget $b$ so the validation vote rate after cooldown does not exceed ten alerts per observed host-day. Frozen thresholds use strict exceedances, and calibration reads no attack labels.

Decisions occur at block end. An eligible candidate starts its node's 24-hour cooldown even if the cap rejects it, but suppressed candidates do not extend cooldown. Remaining candidates are ordered by block, decreasing vote count, then node identifier. The cap admits a candidate only if fewer than ten earlier acceptances remain in the rolling 24-hour interval, expiring exact-day-old acceptances first. Rejected candidates are discarded. Uncapped and capped comparisons use the same candidates and ordering. Delays are measured from campaign start to block end and exclude processing latency.

\begin{figure}[t]\centering\includegraphics[width=\columnwidth]{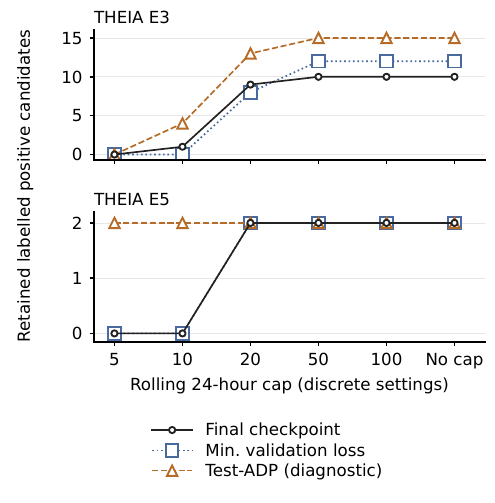}\caption{Corrected-stream capacity sensitivity: retained labelled positive candidates under each rolling 24-hour cap. All six settings per host and checkpoint rule are shown. Test-ADP selection is diagnostic. Each line holds candidates, thresholds, ordering and cooldown fixed. E5's two positives are one episode, and both label-free selectors recover them at cap20. Lines connect discrete tested settings, not estimated performance between them. No independent repetitions or uncertainty intervals are implied.}\label{fig:capacity-sensitivity}\end{figure}

\textbf{Corrected capacity outcomes.} At final checkpoints, THEIA E5 changes from 2 TP / 22 FP without a cap to 0 / 17 at cap ten. THEIA E3 instead retains one of two campaigns, changing from 10 / 35 to 1 / 22. Both label-free E5 checkpoint rules recover both positives at cap twenty, whereas the test-selected rule retains them even at ten. All host, selector and capacity settings appear in Figure~\ref{fig:capacity-sensitivity}. These contrasts measure suppression of labelled candidates, not independently adjudicated malicious activity.

\textbf{Held-out capacity test.} Before inspecting CADETS E5 test scores, we fixed ten corrected Velox members, a four-of-ten vote, final checkpoints, two-hour blocks, 24-hour cooldown and a validation target of ten alerts per observed host-day. The primary cap preserves both campaigns. Table~\ref{tab:e5-capacity} retains every declared sensitivity, including missed campaigns. The five- and twenty-alert sensitivities refit validation calibration, so they compare whole policies rather than isolated caps on the primary candidate stream.

\textbf{Padding and member-vote tests.} The preregistered padding experiment uses ten corrected members, frozen clean thresholds and a four-of-ten vote. A strong-knowledge, score-blind generator adds 1,000 or 5,000 events per campaign in five realizations each. On THEIA E3, adding 5,000 events per campaign raises mean uncapped false alarms from 36 to 3,447 across the five realizations. Injected activity changes admission under the cap while touching original labelled nodes itself. Original-label coverage can therefore increase without better payload detection. Counting injected nodes as attack targets rewards the generator's activity, not necessarily detection of the original attack.

A separate preregistered member-vote test holds FIVEDIRECTIONS and TRACE out from the other comparisons. It fixes five members, minimum-validation-loss checkpoints and system-specific vote thresholds before outcomes. Three pairs fail and the fourth is degenerate, so no pair passes nondegenerately: none improves precision while retaining at least mean-member coverage and detecting an attack. A secondary predictor based on unshared false alerts also fails on its two positive predictions. Post-hoc vote-threshold variation produces isolated successes in opposite directions, not one choice that rescues the held-out test. This negative result is retained, and member voting is not a claimed contribution.

\begin{table*}[tp]\centering\setlength{\tabcolsep}{6pt}\small
\begin{tabular}{lrrrrrl}
\toprule
Setting & $D$ & TP/FP & TP/FP$_c$ & $a\to a_c$ & $C\to C_c$ & $\delta_c$ (min) \\
\midrule
Primary & 1.67 & 3/69 & 2/13 & 43.20$\rightarrow$9.00 & 2$\rightarrow$2 & 149; 345 \\
Target/cap 5 & 1.67 & 2/62 & 0/7 & 38.40$\rightarrow$4.20 & 1$\rightarrow$0 & ---; --- \\
Target/cap 20 & 1.67 & 3/93 & 2/26 & 57.60$\rightarrow$16.80 & 2$\rightarrow$2 & 149; 345 \\
60-minute blocks & 1.71 & 3/69 & 2/13 & 42.15$\rightarrow$8.78 & 2$\rightarrow$2 & 89; 285 \\
Minimum validation loss & 1.67 & 3/68 & 2/13 & 42.60$\rightarrow$9.00 & 2$\rightarrow$2 & 149; 345 \\
\bottomrule
\end{tabular}
\caption{Preregistered CADETS E5 capacity test: the primary policy and all four declared sensitivities. Ten corrected members, $k=4$, 24-hour cooldown. The primary policy uses final checkpoints, 120-minute blocks and a target/cap of ten alerts per day. $D$ is observed host-days, $a$ alerts per observed host-day and $C$ campaigns found out of two. subscript $c$ denotes the rolling cap. $\delta_c$ lists capped campaign-start delays in campaign order, with a dash for a miss. Within each row the cap acts on the same candidate stream. Across rows, the indicated setting changes and calibration is refitted, so these are not isolated cap changes. Block duration also changes observed exposure, so rate differences are not necessarily count differences.}\label{tab:e5-capacity}
\end{table*}

\FloatBarrier
\end{document}